\documentclass[journal]{IEEEtran}

\ifCLASSINFOpdf

\else

\fi

\usepackage{amsmath}
\usepackage{graphicx}
\usepackage{subfigure}
\usepackage{amsmath}

\usepackage{algpseudocode}
\usepackage{cite}
\usepackage{multicol}
\usepackage{multirow}
\usepackage{mathtools}
\usepackage{amssymb}
\usepackage{color}
\usepackage{bm}
\usepackage{indentfirst}
\usepackage{url}
\usepackage{nomencl}
\usepackage{booktabs}
\usepackage{cite}
\usepackage[numbers,sort&compress]{natbib}\usepackage{amsmath}
\usepackage{graphicx}
\usepackage{subfigure}
\usepackage{amsmath}
\usepackage{algorithm}
\usepackage{algpseudocode}
\usepackage{cite}
\usepackage{multicol}
\usepackage{multirow}
\usepackage{mathtools}
\usepackage{amssymb}
\usepackage{color}
\usepackage{bm}
\usepackage{indentfirst}
\usepackage{url}
\usepackage{nomencl}
\usepackage{booktabs}
\usepackage{cite}
\usepackage{algorithm,algpseudocode}
\usepackage[numbers,sort&compress]{natbib}
\usepackage{siunitx}

\begin{document}

\title{Distributed Control of Multi-zone HVAC Systems  Considering Indoor Air Quality}

\author{Yu~Yang,~\IEEEmembership{Student Member,~IEEE,}
	Seshadhri~Srinivasan,~\IEEEmembership{Senior Member,~IEEE,}\\
	Guoqiang~Hu,~\IEEEmembership{Senior Member,~IEEE,}
	and ~Costas~J.~Spanos,~\IEEEmembership{Fellow,~IEEE}
	\thanks{This  work  was  supported  by  the  Republic  of  Singapore’s  National  Research  Foundation  through  a  grant  to  the  Berkeley  Education  Alliance  for  Research  in  Singapore
		(BEARS)  for  the  Singapore-Berkeley  Building  Efficiency  and  Sustainability  in  the
		Tropics  (SinBerBEST)  Program.  BEARS  has  been  established  by  the  University  of  California,  Berkeley  as  a  center  for  intellectual  excellence  in  research  and  education  in
		Singapore.}
	\thanks{Yu Yang and Seshadhri Srinivasan are with SinBerBEST, Berkeley Education 	Alliance for Research in Singapore, Singapore, 138602. E-mail: ( \{yu.yang,  seshadhri.srinivasan\}@bears-berkeley.sg).}
	\thanks{Guoqiang Hu is with the School 	of Electrical and Electronic Engineering, Nanyang Technological University,
		Singapore, 639798. E-mail: (gqhu@ntu.edu.sg).}
	\thanks{Costas J. Spanos is with the Department of Electrical Engineering and 	Computer Sciences, University of California, Berkeley, CA, 94720 USA. Email: (spanos@berkeley.edu).}
}

\maketitle

\begin{abstract}
This paper studies a scalable control method for multi-zone heating, ventilation and air-conditioning  (HVAC) systems to optimize the energy cost  for  maintaining thermal comfort and  indoor air quality (IAQ)  (represented by CO$_2$) simultaneously. This problem is computationally challenging due to the complex system dynamics, various spatial and temporal couplings  as well as multiple control variables to be coordinated. To address the  challenges, we propose a two-level distributed method (TLDM) with a \emph{upper} level and \emph{lower} level control  integrated.  The \emph{upper} level  computes  zone mass flow rates for maintaining zone thermal comfort with minimal energy cost, and  then the \emph{lower} level  strategically regulates zone mass flow rates  and the ventilation rate to achieve IAQ while preserving the \emph{near} energy saving performance of \emph{upper} level.  As both the \emph{upper} and \emph{lower} level computation are deployed in a distributed manner, the proposed method is scalable and computationally efficient.  The near-optimal performance of the method in energy cost saving  is  demonstrated through comparison with  the  centralized method. In addition, the comparisons with the existing distributed method  show  that our method can provide IAQ with only  little increase of  energy cost while the latter fails.
Moreover, we demonstrate our method outperforms the demand controlled ventilation strategies (DCVs) for IAQ management  with  about  $8$-$10\%$ energy cost reduction.  
\end{abstract}

\renewcommand\abstractname{Note to Practitioners}
\begin{abstract} 
The high portion of building energy consumption  has motivated the energy saving for HVAC systems. Concurrently, the living standards for indoor environment are rising among the occupants. Nevertheless,  the status quo on improving building energy efficiency has mostly focused on  maintaining thermal comfort (such as temperature),  and  the indoor air quality (IAQ) (usually represented by CO$_2$ level)  has been 
	 seldom incorporated. 

In   our previous work with the similar setting,  we observed that the CO$_2$ levels will surge beyond tolerance during the high occupancy periods if only thermal comfort is considered for HVAC control.  This deduces the IAQ and thermal comfort should be jointly considered while pursuing  the energy cost saving target and thus studied in this paper. This task is computationally cumbersome due to the complex system dynamics (thermal and CO$_2$) and tight correlations among the different control components (variable air volume and fresh air damper). To cope with these challenges, this work develops a two-level distributed computation paradigm for HVAC systems based on problem structures.
Specifically, the \emph{upper} level control first calculates zone mass flow rates for maintaining  comfortable zone temperature with minimal energy cost and then  the \emph{lower} level strategically regulates the computed zone mass flow rates as well as  ventilation rate  to satisfy IAQ while preserving the \emph{near} energy saving performance of the \emph{upper} level control.  As both the \emph{upper} and \emph{lower} level  calculation can be implemented in a distributed manner,  the proposed method is scalable  to  large multi-zone deployment.  The method's performance both in maintaining comfort (i.e., thermal comfort and IAQ) and energy cost saving is demonstrated via simulations in comparisons with the centralized method,  the distributed token-based scheduling strategy and the demand controlled ventilation strategies. 
\end{abstract}


%
\IEEEpeerreviewmaketitle

\section{Introduction}
The heating, ventilation and air-conditioning (HVAC) systems account for 40\%-50\% of  building energy consumption for maintaining comfortable indoor environment \cite{ku2015automatic}. Aware of the high portion of energy consumption, the proposition to improve energy efficiency for HVAC systems has stimulated widespread attention~\cite{mirinejad2012review, afram2014theory}. A plenty of works have demonstrated substantial energy can be saved by deploying advanced HVAC control strategies (see \cite{kelman2011bilinear, radhakrishnan2016token},  for examples). However,  regarding the human sensation, the main focus has been placed  on  thermal comfort which is usually indicated by temperature and humidity \cite{yang2019stochastic, xu2017pmv},  the indoor air quality (IAQ), such as the carbon dioxide (CO$_2$) concentration has been seldom studied.

IAQ is closely related to the mechanical ventilation rate (i.e., fresh air infusion) of HVAC systems, especially nowadays where most  buildings are constructed in closed envelope for energy saving concerns.   IAQ has emerged as  a critical issue along developing building automation.
 On one hand, the   living standards and awareness of health  are rising among the occupants;  on the other hand,  IAQ is closely related to  occupant well-being and working productivity~\cite{dasi2010line, li2018novel}. 
A study conducted by 
 the National Institute of Environmental Health Science (NIEHS) in 2015 has demonstrated the  impact of IAQ  on cognitive abilities~\cite{Cognitive}. 
Significantly, the findings show that the strategic skills of occupants  reduce to 20\% with an indoor CO$_2$ concentration 1,400~ppm compared to the normal outdoor level of 400~ppm.
Straightforwardly,  for energy-efficient HVAC control, as the thermal comfort only relates to zone mas flow rates,  the fresh air ventilation of HVAC systems is usually  suppressed to the lower boundary to minimize the cooling load.  However,  for IAQ management,  sufficient  ventilation is required to dilute the indoor air. 
This implies considering thermal comfort alone is not enough for developing energy-efficient HVAC control strategies. As in such setting, the energy saving target may be achieved at the expense of awful IAQ.  Specifically, we can imagine that  the CO$_2$ concentration will accumulate and surge beyond  tolerance  during the time periods with high occupancy though the temperature is maintained comfortable if sufficient fresh air ventilation is not provided by the HVAC systems.


Notably, increasing focus has been shifted towards  IAQ management for HVAC systems both from industries and research communities in recent years. Some industries have informed that the thermostats will integrate  CO2 sensors inside for monitoring IAQ in foreseeable future \cite{CO2sensor}.  Within the research communities, some pioneering works  have jointly considered  IAQ and thermal comfort while studying energy-efficient HVAC control (see \cite{parisio2013scenario, parisio2014implementation, yu2018energy}, for examples).  Nevertheless,  this  problem  hasn't been well addressed yet  as
\emph{i)} the related works are still fairly limited (see the references therein); and
\emph{ii)} most of them are for single-zone cases \cite{parisio2013scenario, parisio2014implementation} or not scalable to multi-zone commercial buildings due to the centralized computation framework \cite{yu2018energy}. 
The  computational challenges of the complex problem remain   to be addressed: 
\begin{itemize}
	\item[{\em{i)}}] Multiple control variables need to be coordinated for achieving the multiple objectives.  Both the  ventilation  rate and zone mass flow rates require  to be coordinated  as to optimize  energy efficiency for achieving  thermal comfort and IAQ simultaneously. Particularly, a good IAQ  requires  high ventilation rate and zone mass flow rates. However, this may induce lower temperature bound violations  and high energy cost.
	\item[{\em{ii)}}] The  problem is non-linear and non-convex due to the complex system dynamics.  Both the energy cost and system dynamics  (i.e., temperature and CO$_2$) are non-linear w.r.t the control inputs:  zone mass flow rates and ventilation rate.
	\item[{\em{iii)}}] There exist various spatial and temporal couplings. The optimal operation of  HVAC systems should consider zone temperature and CO$_2$ inertia, which correspond to substantial temporal constraints. Besides, the inter-zone heat transfer and  the recirculated air to AHU induce tight spatial couplings. 
	
\end{itemize}

\subsection{Contributions}
Motivated by the literature, this paper studies the control of  multi-zone commercial HVAC systems to optimize the energy cost for  maintaining thermal comfort and IAQ simultaneously.  We make the following contributions to overcome the computational challenges.
\begin{itemize}
\item[\em{(C1)}] We propose a  two-level control method integrated with  the \emph{upper} level control (ULC) and the \emph{lower} level control (LLC) by exploiting the problem structures.  

\item[\em{(C2)}] While the ULC adopts an existing distributed method, we develop a distributed method for the LLC to achieve scalability and computation efficiency.

\item[\em{(C3)}] We demonstrate the method's performance  in energy cost saving,  human comfort (i.e., temperature and CO$_2$) as well as computation efficiency through simulations. 
\end{itemize}

Our two-level structure is motivated by the independent zone temperature and zone CO$_2$ dynamics (but both subject to the control inputs),  which makes it possible to tackle the two comfort indexes successively. 
 Therefore, we decompose the problem into two levels: the ULC first computes the optimal zone mass flow rates to for maintaining zone thermal comfort with minimal energy cost. Successively,  the LLC strategically regulates the computed zone mass flow rates from ULC and ventilation rate to achieve the desirable IAQ.  Such two-level  paradigm makes it computationally tractable to achieve the two comfort indexes simultaneously  while preserving the \emph{near-optimal} energy cost.  Moreover,  the two-level structure favors computation efficiency   as the LLC is only invoked  when the  CO$_2$  upper bounds are  to be violated. 
Particularly,  as the ULC on thermal comfort has been comprehensively studied in our previous work \cite{yu2018TASE} and thus adopted, we place our main focus  on bringing in the two-level structure and the  LLC.

		 

The remainder of this paper is structured as below. 
In Section II, we review the related works. 
In Section III,  we present the problem formulation.  In Section IV, we discuss the two-level distributed method.
In Section V,   we evaluate the performance of the method via simulations.
In Section VI, we briefly conclude this paper.

\section{Related Works}

Developing advanced control for  HVAC systems  to improve energy efficiency has been an edge issue along  building automation over the past decades.  Various methods have been studied and some comprehensive review can refer to  \cite{mirinejad2012review, afram2014theory}.   
Significantly, these existing works have proved that substantial energy can be saved by deploying advanced HVAC control while not compromising human comfort. From the computation structure standpoint,   these methods can  be categorized into  centralized methods~\cite{kelman2011bilinear} and decentralized methods~\cite{radhakrishnan2016token}.  Centralized approaches are usually developed for single-room/zone cases   and not amenable to commercial buildings due to the substantial computation burden (see \cite{yang2019stochastic, xu2017pmv}, for examples).  By contrast, 
a number of decentralized approaches have been developed to address the computation challenges  of commercial HVAC systems \cite{radhakrishnan2016token, yu2018TASE, mei2019distributed}. 

Nevertheless, one critical issue to be noticed is that most of these works have focused on thermal comfort (i.e., temperature, humidity, etc.)  and  lack  IAQ management while pursuing energy cost savings~(see  \cite{okaeme2017passivity, okaeme2016comfort} and the references therein).  In this backdrop, the IAQ may be awful and beyond tolerance though the temperature is maintained comfortable. For example, the pollutants and particles, especially CO$_2$ concentration  accumulate over the time if sufficient ventilation (i.e, outdoor fresh air) is not provided.
This is easy to understand as  high proportion of recirculated air  is preferred to reduce the cooling load.  
In the literature, the IAQ management is mostly addressed by some simple ventilation rules referred to  the  demand controlled ventilation  strategies (DCVs). 
Such methods could be CO$_2$-based \cite{nassif2005ventilation, nielsen2010energy,hesaraki2015influence} or occupancy-based \cite{wang2013intelligent, nassif2012robust, marinovsensor}.  The  main ideas  are adjusting the amounts of fresh air infusion   based on the detected instantaneous CO$_2$ concentration or occupancy. 
Whereas  for multi-zone commercial buildings, the DCVs  tend to cause  over-ventilation  or under-ventilation   due to zone CO$_2$ or occupancy variations ~\cite{shan2012development}. 
Another drawback is that the ventilation regulation  for IAQ management  are not coordinated with the thermal comfort, which will jeopardize the energy saving objective of HVAC systems. 

It's imperative to jointly consider both thermal comfort and IAQ  simultaneously   both for achieving   HVAC energy cost savings and maintaining human comfort.   Such awareness has motivated  the joint management of thermal comfort and IAQ  for  single-zone cases \cite{parisio2013scenario, parisio2013randomized, parisio2014implementation}. Particularly,  most of these methods  depend on simplified linear models to capture  system dynamics. Therefore,  the methods and  modelings are generally not amenable to  multi-zone commercial buildings. 
As a scarce exception,  \cite{yu2018energy}  studied  both thermal comfort and IAQ  management for commercial HVAC systems based on Lyapunov optimization.  
The performance of the method both in  maintaining human comfort (i.e., thermal comfort and IAQ)  and saving energy cost was demonstrated on a $4$-zone case study. However, the computation tractability  for  larger scale applications remains to be addressed and motivates this work.

\section{Problem Formulation}

\subsection{HVAC  Systems for Commercial Buildings}

\begin{figure}
	\centering
	\includegraphics[width=3.4 in]{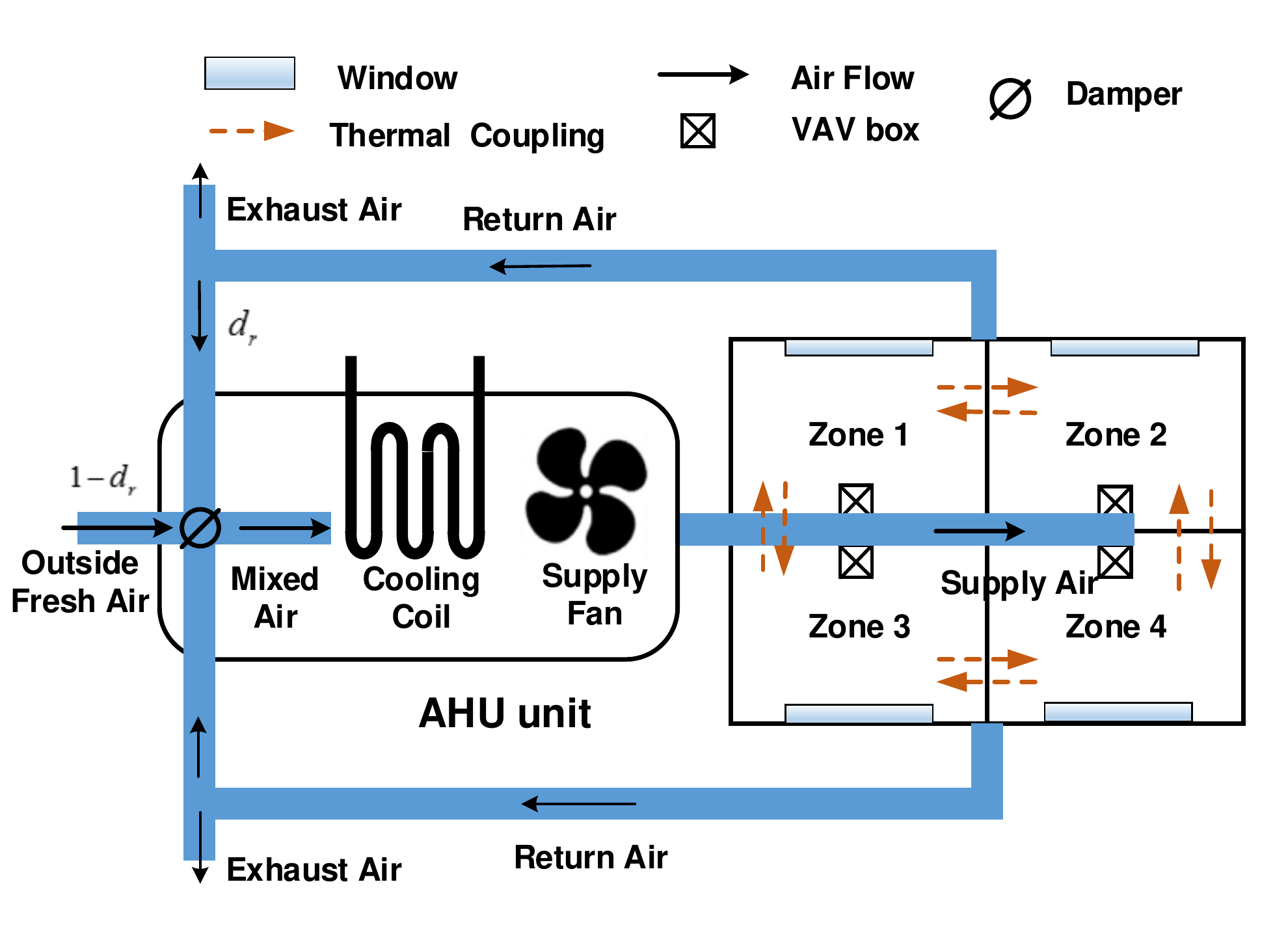}\\
	\caption{The schematic of the HVAC system for multi-zone buildings.}\label{system architecture}
\end{figure}

The configuration for a  typical  commercial HVAC system  is shown in Fig. \ref{system architecture}, which mainly consists of an  Air Handling Unit (AHU),  Variable Air Volume (VAV) boxes  and a  chiller  system (not shown here). 
 The  AHU is  basically  integrated with a damper, a cooling/heating coil and a supply fan. 
The heating/cooling coil  cools down/heats up the mixed air (the mixture of outside fresh air and  inside recirculated air) to the set-point temperature.  
The damper regulates  the ventilation rates: the fraction of  recirculated air $d_{r}$ or the fraction of outdoor fresh air $1-d_r$.  Therefore,  a smaller $d_{r}$ specifies higher proportion of fresh air infusion but induces higher cooling demand for the HVAC system, and vice versa.  The supply fan is for driving the circulation of air within the building duct network.
As another main component,  the VAV boxes are connected to the zones, each of which  
consists of a damper and a heating coil.  The damper regulates the zone mass flow rate  and the heating coil reheats the supply air if necessary (not discussed in this paper).  
The chiller  system is usually constituted by a chiller pump, water tank and the chiller,  which provides  continuous chilled water to the cooling coils of  AHU.  Except for the chiller,  the chiller pump is also partially  responsible for the HVAC's energy consumption  for circulating the water between the water tank and the chiller.   This paper  studies HVAC systems with  constant water flow system \cite{ConstantWaterFlow} where the chiller pump's energy consumption can be regarded as fixed and thus not explicitly discussed.  Besides, without loss of generality, this paper studies  the cooling mode. 
More details of the  HVAC systems  can refer to  \cite{kelman2011bilinear, zhang2017decentralized}.

As aforementioned, we seek to minimize the energy cost for  maintaining
both zone thermal comfort and IAQ simultaneously. 
In such setting,   both zone air flow rates  and ventilation rate  ($d_{r}$) need  to be coordinated.  
The problem is studied in a discrete-time framework  on a daily basis with  $\Delta_k= 30$ \si{\minute}'s sampling and computing epoch. To account for the multi-stage control under
 uncertainties (e.g., weather, occupancy, etc.),  we deploy the  model predictive control (MPC) framework: the control inputs at each executed epoch are  computed based on the predicted information over a  look-ahead planning horizon $H=10$ (5\si{\hour}). This process is repeated with the time evolving until the end of the optimization horizon $\mathcal{T}$.

\subsection{Zone Thermal  Dynamics}
We consider a commercial building with $I$ zones indexed by $\mathcal{I}=\{1, 2, \cdots, I\}$. 
At each computing epoch, the zone thermal dynamics over the planning horizon $\mathcal{H}=\{0, 1, \cdots, H-1\}$ are captured by  a Resistance-Capacitance (RC) network  \cite{lin2012issues, maasoumy2011model}:

\vspace{-3mm}
{\small{ 
\begin{equation} \label{zone temperature dynamics}
\begin{split}
&C^{\rho}_i (T_i(k+1)\!-\!T_i(k))\!=\!\sum_{ \mathclap{ j\in \mathcal{N}_i}}\! \frac{T_j(k)\!-\!T_i(k) }{R_{ij}}  \Delta_k\!+\!\frac{T_o(k)\!-\!T_i(k) }{R_{oi}} \!\Delta_k\\
&\quad +c_p m_i^{z}(k)(T_c-T_i(k) )\Delta_k+Q_i(k) \Delta_k, ~\forall i \in \mathcal{I},  k \in \mathcal{H}.
\end{split}
\end{equation}}}
where $k \in \mathcal{H}$ and $i ,j \in \mathcal{N}$ denote the time and zone indices. $C^\rho_i$ is the zone air heat capacity.  $T_i(k)$,  $T_o(k)$ and $T_c$ denote the zone temperature,  outdoor temperature and  the set-point temperature of AHU, respectively. 
$R_{ij}$ ($R_{ji}$) denotes the adjacent zone thermal resistance. Particularly, $R_{oi}$ denotes the thermal resistance between zone $i$ and the outside, We use $\mathcal{N}_i$ to indicate the set of spatially adjacent zones to zone $i$. 
$c_p$ is the specific heat of the air. $m^{z}_i(k)$ indicates the zone mass flow rate.
$Q_i(k)$ captures zone internal  heat gains from the occupants and electrical equipment, which can estimated by the occupancy \cite{korolija2011influence}.   

We translate (\ref{zone temperature dynamics}) into a standard form:
{\small{
\begin{equation} \label{zone temperature dynamics2}
\begin{split}
&T_i(k+1)=A_{ii}T_i(k)+\sum_{j\in \mathcal{N}_i} A_{ij} T_j(k)\\
&\quad \quad +C_{ii} m^{z}_i(k) (T_i(k)-T_c)+D_i(k),  ~\forall i \in \mathcal{I},  k \in \mathcal{H}.
\end{split}
\end{equation}
where $A_{ii}=1\!-\!(\sum_{j\in \mathcal{N}_i} \frac{\Delta_k}{R_{ij}C^\rho_i}+\frac{\Delta_k}{C^{\rho}_i  R_{oi}})$, $A_{ij}\!=\!\frac{\Delta_k}{C^\rho_i R_{ij}}$,
$C_{ii}\!=\!-\frac{\Delta_k \cdot  c_p}{C^{\rho}_i} $, and $D_{i}(k)=\frac{\Delta_k T_o(k)}{C^{\rho}_i R_{oi}}+\frac{\Delta_k \cdot Q_i(k)}{C^{\rho}_i}$. 
}}

Similar to \cite{parisio2013scenario, yu2018energy}, this paper uses CO$_2$ concentration as an IAQ indicator. 
The  zone CO$_2$ dynamics  for multi-zone  commercial buildings can be described by \cite{yu2018energy}

\vspace{-4mm}
{\small{
\begin{equation} \label{CO2 dynamics}
\begin{split}
m_i(C_i(k+1)&-C_i(k))=  N_i(k)  C_g \Delta_k \\
  & +m_i^{z}(k) (C_{z}(k)-C_i(k)) \Delta_k ~\forall i \in \mathcal{I},  k \in \mathcal{H}\\
\end{split}
\end{equation}}}
where $m_i$ denotes zone air mass. $C_i(k)$ (in ppm) denotes zone CO$_2$ concentration. 
We suppose the occupants are  the main source of CO$_2$ generation and thus the CO$_2$ accumulation  can be estimated by  the average CO$_2$ generation rate per person $C_g$  (\si{\gram\per\hour}) multiplied by the occupancy $N_i(k)$  as indicated in the first term of  right-hand side.
$C_z(k)$ denotes the  CO$_2$ concentration of  supply air, which can be estimated by

\vspace{-3mm}
{\small{
\begin{equation} \label{CO2 dynamics of supply air}
\begin{split}
&C_z(k)=\big(1-d_{r}(k) \big) C_o(k)+d_{r}(k)C_{\textrm{m}}(k),\\
&\textrm{with}~C_{\textrm{m}}(k)=\frac{\sum_{i \in \mathcal{I}} m_i^{z}(k) C_i(k)}{\sum_{i \in \mathcal{I}} m_i^{z}(k)}, ~\forall  k \in \mathcal{H}
\end{split}
\end{equation}}}
where $d_{r} (k) \in [0, 1]$ denotes the fraction of return air  delivered to AHU (i.e., the ventilation rate to be controlled).  $C_{m}(k)$  characterizes the average CO$_2$ concentration of the return air from all  zones.

From (\ref{CO2 dynamics}) and (\ref{CO2 dynamics of supply air}), we note that the zone  CO$_2$ dynamics are  non-linear and  fully coupled through the recirculated air.

\subsection{Decision Variables}

Our objective is to minimize the HVAC energy cost for maintaining thermal comfort and IAQ.  Therefore,  our decision variables can be divided into  control variables and state variables. The control variables are the control inputs of the HVAC system that affect indoor condition, which include the ventilation rate $d_{r}(k)$ and  zone mass flow rate $m_i^{z}(k)$.
Our state variables are zone temperature $T_i(k)$ and  zone CO$_2$ concentration $C_i(k)$, which indicate human comfort and requires to be maintained.


\subsection{Objective Function}
The energy consumption of the  HVAC system is mainly incurred  by the cooling coil $P_f(k)$ and supply fan $P_f(k)$ within AHU, i.e., 
\begin{equation} \label{AHU cooling power2}
\begin{split}
&P_c(k)=c_p \eta (1-d_{r}(k))\sum_{i \in \mathcal{I}} m_i^{z}(k)(T_o(k)-T_c)\!\\
     &\quad \quad \quad +c_p \eta d_{r}(k) \sum_{i \in \mathcal{I}} m_i^{z}(k)( T_i(k)-T_c )\\
   &  P_f(k)=\kappa_f (\sum_{i \in \mathcal{I}} m_i^{z}(k))^2
\end{split}
\end{equation}
where $\eta$ is  the  reciprocal  of the coefficient  of  performance  (COP)  of  the  chiller, which captures the ratio of provided cooling to the total consumed electrical power. 

Considering the energy consumption is not easy to inspect in practice, we selected the total energy cost  charged by the  electricity price  $c_k$ (s\$\si{\per{\kilo\watt}})  on a daily basis as our  objective:
 \begin{equation}
 J=\sum_{k \in \mathcal{H}}  c_k \big(  P_c(k) +P_f(k)\big) \Delta_k
 \end{equation}

\subsection{System Constraints}

The HVAC operation should respect the thermal comfort and IAQ requirements,  which are represented by zone temperature bounds  \cite{radhakrishnan2016token}  and  CO$_2$ bounds  \cite{yu2018energy} in this paper:
\begin{equation} \label{temperature ranges}
\begin{split}
{T}^{\min}_i & \leq T_i(k) \leq {T}^{\max}_i, \quad \forall i \in \mathcal{I}, ~k \in \mathcal{H}.
\end{split}
\end{equation}
\vspace{-6mm}
\begin{equation} \label{CO2 ranges}
\begin{split}
C_i(k) \leq  C^{\max}_{i},  \quad \forall i \in \mathcal{I}, ~k \in \mathcal{H}.
\end{split}
\end{equation}
where ${T}^{\min}_i$ and ${T}^{\max}_i$ characterize the comfortable  zone temperature ranges. 
$C_i^{\max}$ denotes the desirable  zone CO$_2$ upper bound. 
Particularly, the formulation  can account for personalized human comfort  by differentiating  ${T}^{\min}_i$ , ${T}^{\max}_i$  and $C_i^{\max}$ w.r.t the zones. 

Additionally,   the HVAC operation  must abide by the physical limits of VAV boxes and AHU: 
\begin{equation} \label{zone air flow rate bound}
\begin{split}
&{m}^{z, \min}_i \leq m^{z}_i(k) \leq m_i^{z, \max}, \quad \forall i \in \mathcal{I}, ~k \in \mathcal{H}\\
\end{split}
\end{equation}
\vspace{-5mm}
\begin{equation} \label{total air flow rate bound}
\begin{split}
&\sum_{i \in \mathcal{I}} m_i^{z}(k) \leq {m}^{\max}, ~~~\forall k \in \mathcal{H} \quad \quad \quad \quad~~\\
\end{split}
\end{equation}
where $m^{z, \min}_i$ and $m^{z, \max}_i$ denote the operation range of VAV boxes. 
$m^{\max}$ denotes the supply capacity of  AHU. 

Similarly, the damper of AHU for regulating ventilation rate is usually restricted to the operation range characterized by  $d^{\min}_r$ and ${d}^{\max}_r$ to reduce wear and tear: 
\begin{equation} \label{ventilation rate}
\begin{split}
d^{\min}_r \leq d_r(k) \leq  d^{\max}_r, ~~\forall k \in \mathcal{H}. 
\end{split}
\end{equation}

\vspace{-6mm}
\subsection{The Problem}
Overall, the optimization problem at each computing epoch can be summarized as \eqref{obj}.
\begin{align}
\label{obj} &\min_{\bm{m}^{z}_i, \bm{T}_i, \bm{C}_i, i \in \mathcal{I}.  \bm{d}_{r} } J \tag{$\mathcal{P}$ }\\
s.t.~ (\ref{zone temperature dynamics2})\!&-\!(\ref{CO2 dynamics of supply air}), (\ref{temperature ranges})\!-\!(\ref{CO2 ranges}), ~\eqref{zone air flow rate bound}\!-\!\eqref{total air flow rate bound}, ~\eqref{ventilation rate}. \quad \quad  \notag
\end{align}
where we have $\bm{m}^{z}_i=[m^{z}_i(k)]_{k \in \mathcal{H}}$, $\bm{T}_i=[T_i(k)]_{k \in \mathcal{H}}$, $\bm{C}_i=[C_i(k)]_{k\in\mathcal{H}}$ ($\forall i \in \mathcal{I}$) and $\bm{d}_r=[d_r(k)]_{k \in \mathcal{H}}$ denoting the concatenating  decision variables over the computing epoch. 

It's clear to see that problem \eqref{obj} is  non-linear and non-convex. 
The non-linearity and non-convexity both arise from the objective function and the constraints. 
Moreover, the  substantial spatially and temporally coupled constraints imposed by zone temperature and zone CO$_2$ makes it computationally intractable even for moderate commercial HVAC systems with centralized methods.
Indeed,  this motivates our  two-level  distributed  method  to be discussed.




\section{Two-Level Distributed Method}
\begin{figure}
			\setlength{\abovecaptionskip}{-6pt}
	\setlength{\belowcaptionskip}{6pt}
	\centering
	\includegraphics[width=3.2 in]{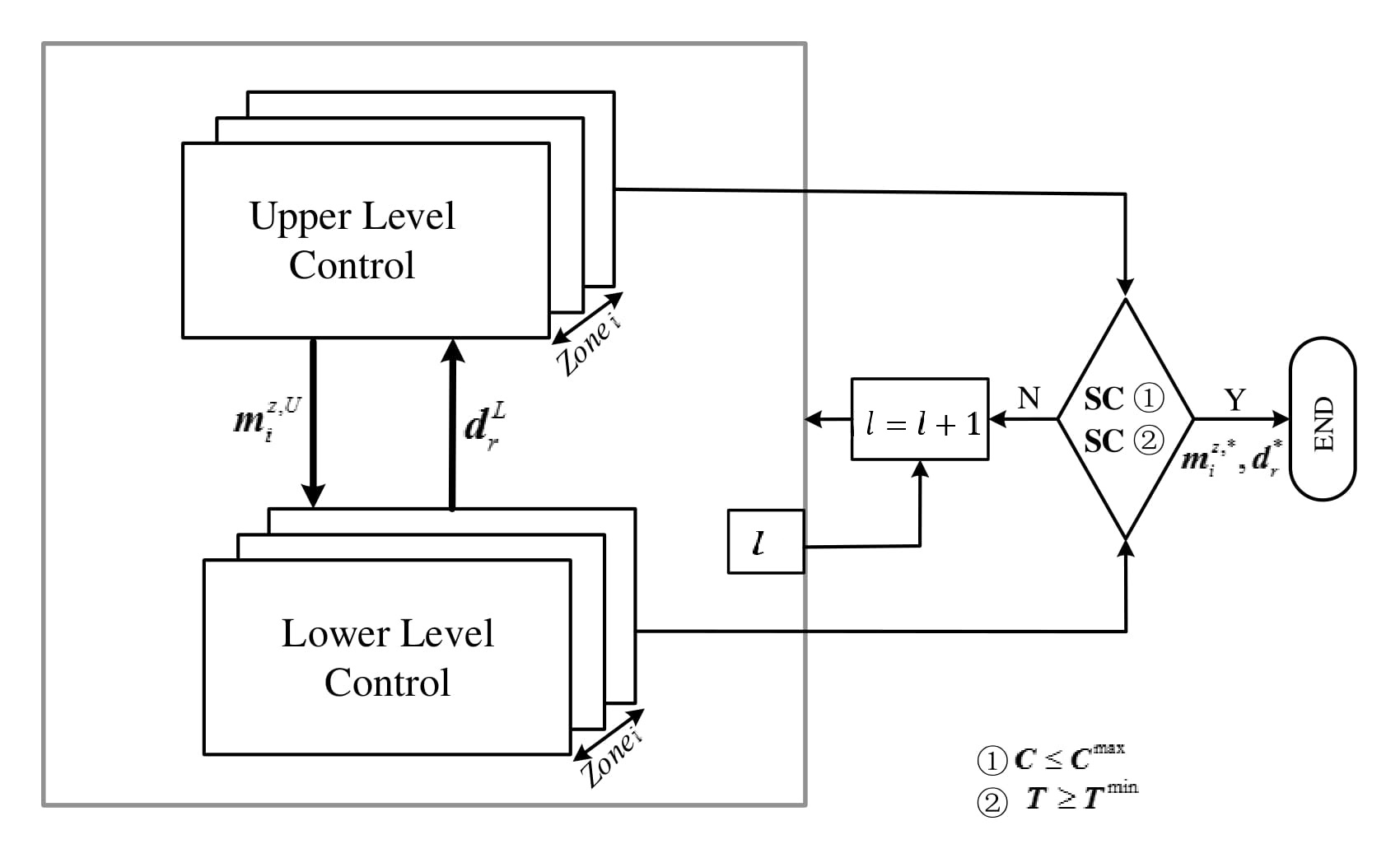}\\
	\caption{The framework of two-level distributed method.}\label{hierarchical approach} \label{framework}
\end{figure}

To address the computational challenges of problem \eqref{obj},  we propose a  two-level distributed method (TLDM) based on  the  problem structures: the zone temperature and CO$_2$ dynamics are independent (but both subject to the control inputs). We divide the problem into two levels:  the \emph{upper} level control (ULC) and the \emph{lower} level control (LLC). 
The  ULC  is  mainly responsible for thermal comfort  and the  LLC corresponds to IAQ management.
The underlying idea of  the two-level structure is  to address the  two comfort indexes successively  
 while preserving the \emph{near-optimal} energy saving performance.
Moreover, such two-level structure favors computation efficiency as the LLC requires to be invoked only if the CO$_2$  is to be violated. 
Both  the ULC and LLC use distributed implementation  to achieve scalability and computation efficiency.     
 We present the holistic framework of the TLDM  in Fig.~\ref{framework}.  
 Particularly, the ULC first computes the zone flow rates $\bm{m}^{z, U}$ for maintaining thermal comfort with minimal energy cost  and then the LLC strategically regulates the ventilation rate $\bm{d}^L_r$ to achieve IAQ while preserving the \emph{near} energy cost of ULC. The ULC and LLC alternate and communicate with each other.   The implementation of   ULC and LLC are as below.
 
\subsection{The Upper Level  Control (ULC)}

The ULC focuses on maintaining zone thermal comfort with minimal  energy cost.
Considering  that \emph{i)} zone temperature is only affected  by   zone mass flow rate $m^z_i(k)$ and \emph{ii)} the HVAC's energy cost is non-decreasing w.r.t. the ventilation rate $d_r(k)$, 
we define the ULC problem as 
\begin{align}
& \quad \quad   \min_{  \bm{m}^{z}_i, \bm{T}_i, i \in \mathcal{I}.}  J^{{U}} \notag\\
&\label{P2} s.t.~~ (\ref{zone temperature dynamics2}), (\ref{temperature ranges}), (\ref{zone air flow rate bound})\!\!-\!\!(\ref{total air flow rate bound}). \quad \quad  \tag{$\mathcal{P}_U$}\\
&\quad \quad d_r(k)=d^L_r(k), k \in \mathcal{H}. \notag
\end{align}
where we have $J^{{U}}=J$ and $\bm{d}^L_r = [d^{L}_r(k)]_{k \in \mathcal{K}}$. 
The intuitive interpretation of  ULC is  to  identify the zone mass flow rates for maintaining thermal comfort with minimal energy cost under the specified ventilation rate $d^L_r(k)$ from LLC (we can start with ${d}^{\max}_r$ ($k \in \mathcal{H}$)).
For  problem \eqref{P2}, there already exists a number of distributed methods~\cite{yu2018TASE, radhakrishnan2016token}. 
We adopt our previously proposed distributed method \cite{yu2018TASE} and  place our main attention  on  the implementation of  LLC.

\subsection{The Lower Level  Control (LLC)}
As indicated in Fig.  \ref{framework}, the LLC  only needs to be  invoked when the zone CO$_2$ concentrations violate the upper bounds (i.e., $C_i(k) \geq C^{\max}_i$). 
Clearly, both increasing zone mass flow rates and ventilation rate (decrease $d_{r}(k)$) can dilute zone  CO$_2$ thus improve IAQ.  However, the latter is generally more expensive as it increase the cooling load of all zones. 
With the concern to preserve the \emph{near} energy cost computed in ULC, our LLC adopts a  \emph{two-phase} method to achieve IAQ.  
We first seek to regulate zone mass flow rates to satisfy the user-defined zone CO$_2$  bounds.
To preserve the \emph{near} energy cost of ULC,  we select the deviation of zone mass flow rates relative to the ULC calculations as the objective. We have the following problem for the first phase of LLC:
\begin{align}
&  \label{P4}  \min_{  \bm{m}^{z}_i,  \bm{C}_i, i \in \mathcal{I}. }  J^{L}=\sum_{ k \in \mathcal{H}}   \sum_{i \in \mathcal{I}}  \big(m^{z}_i(k)-m^{z, U}_i(k) \big)^2   \tag{$\mathcal{P}_L$}\\
&\label{eq:11}~~~s.t.~~ m^{z, U}_i(k) \leq m^{z}_i (k) \leq {m}^{z, \max}_i , \forall i \in \mathcal{I}. \\
&\quad \quad \quad \quad \quad \quad~(\ref{CO2 dynamics})\!-\!(\ref{CO2 dynamics of supply air}), (\ref{CO2 ranges}),(\ref{total air flow rate bound}). \quad \quad  \notag 
\end{align}
where constraints \eqref{CO2 dynamics}-\eqref{CO2 dynamics of supply air}  and \eqref{CO2 ranges} capture zone CO2 dynamics and bounds. Constraints (\ref{total air flow rate bound}) denotes the supply capacity of AHU. Particularly, constraints (12) model the zone mass flow rate lower bounds for maintaining thermal comfort, which is obtained from the ULC.  

 We suppose to achieve thermal comfort and IAQ simultaneously by solving problem \eqref{P4}, however things may not always happen in that way.
  We note  that  only the upper comfortable zone temperature bounds  can be maintained but not necessarily the lower ones by \eqref{eq:11}.  Therefore, it may occur that the desirable CO$_2$ is achieved by solving problem \eqref{P4} but the zone temperature drops beyond the lower bounds due to the substantial increase of zone mass flow rates.    Actually, this means regulating zone mass flow rates alone is  not viable to achieve the two comfort indexes simultaneously, and the  second  phase of LLC should be invoked to increase the ventilation  rate (decrease $\bm{d}_r$).    As illustrated in Fig. \ref{framework}, 
such two phases alternate  until  we achieve the two comfort indexes.
 Suppose the \emph{two-phase} structure gets well across, the  remaining problem is about the implementation. 
 
 \emph{First phase of LLC}: Clearly, for the first phase, we require to solve the non-linear and non-convex problem \textbf{(\ref{P4})}, which is a nontrivial task.  
We note that the zone CO$_2$ dynamics are fully coupled though the recirculated air, posing the primary challenge to develop a distributed method.  
To handle this,  we introduce a learning framework to estimate  CO$_2$ concentration for the supply air $C_z(k)$.
Therefore,  the  procedure to solve problem \eqref{P4} contains two successive steps:  first estimate $C_z(k)$ and then solve problem of \eqref{P4} with the estimated $C_z(k)$. 
Specifically, with the estimated  $\bm{C}_z(k)$,  we have the decoupled zone CO$_2$ dynamics:
\begin{equation}  \label{CO2 dynamics2}
\begin{split}
C_i(k+1)=C_i(k)+&E_{i}(k) m^{z}_i(k) \\
  &+F_i(k) m^{z}_i(k) C_i(k)+G_i(k)
\end{split}
\end{equation}
where $E_{i}(k)={C_z(k)\Delta_k}/{m_i}$, $F_i(k)={-\Delta_k}/{m_i}$ and $G_i(k)={ N_i(k) C_g \Delta_k}/{m_i}$ are now constant parameters.

We note that the zone CO$_2$ dynamics \eqref{CO2 dynamics2} are  bilinear w.r.t the  zone mass flow rate $m^{z}_i(k)$ and  zone CO$_2$ $C_i(k)$. 
To address the non-linearity, we introduce some auxiliary decision variables $Z_i(k)=m^{z}_i(k) C_i(k)$ and 
use  the \emph{McCormick envelopes}  \cite{mccormick1976computability} to relax the bilinear terms: 
\begin{align}
&\label{13a} Z_i(0)=m^{z}_i(0) C_i(0).  \tag{13a}\\
&\label{13b} Z_i(k) \geq  m^{z, \min}_i C_i(k)+m^{z}_i(k) C^{\min}_i- m^{z, \min}_i C_i^{\min},  \tag{13b}\\
&\label{13c} Z_i(k) \!\geq\! m^{z, \max}_i C_i(k)\!+\!m^{z}_i(k) {C}^{\max}_i\!-\!{m}^{z, \max}_i {C}^{\max}_i, \tag{13c}\\
&\label{13d}  Z_i(k) \leq   m^{z}_i(k){C}^{\max}_i+{m}^{z, \min}_i C_i(k)- {m}^{z, \min}_i {C}^{\max}_i,  \tag{13d}\\
& \label{13e} Z_i(k) \leq {m}^{z, \max}_iC_i(k)+m^{z}_i(k) {C}^{\min}_i-{m}^{z, \max}_i {C}^{\min}_i, \tag{13e}	\\	                              
&\quad \quad \quad \quad \quad \forall k \in \mathcal{H}\setminus\{0\}.   
\end{align}
where $C_i(0)$ denotes the measured  zone CO$_2$ at the beginning of current computing epoch. 

By invoking  the auxiliary decision variables $Z_i(k)$,  we have the following  relaxed convex problem for problem \eqref{P4}:
\begin{subequations}
\begin{align}
& \label{P5} \min_{  \bm{m}^{z}_i,  \bm{C}_i, \bm{Z}_i, \forall i \in \mathcal{I}. }  J^{L}=\sum_{k \in \mathcal{H}} \sum_{i \in \mathcal{I}}   (m^{z}_i (k)-m^{z, U}_i(k))^2  \tag{$\mathcal{P}^{'}_L$}\\
&s.t.~~ C_i(k+1)=C_i(k)+E_{i}(k) m^{z}_i(k) +F_i(k) Z_i(k) \notag\\
&  \label{(12a)} \quad \quad \quad  \quad \quad \quad+G_i(k), ~~\forall i \in \mathcal{I}, k \in \mathcal{H}. \tag{14a}   \\
& \quad \quad \quad \quad \quad  (\ref{CO2 ranges}), (\ref{total air flow rate bound}),   \eqref{eq:11}, (\ref{13a})-(\ref{13e}). \quad \quad  \notag 
\end{align}
\end{subequations}

We note problem \eqref{P5} is  characterized by  \emph{i)} a decomposable  objective function  w.r.t. the zones and  \emph{ii)} coupled linear constraints. This problem  
can be efficiently tackled  by the Accelerated Distributed Augmented Lagrangian (ADAL) method  \cite{chatzipanagiotis2015augmented}, which may be more clearly seen by  recasting  it  into a standard form:
\begin{align}
\min_{\bm{x}_i, i\in \mathcal{I}} & J^{L}=\sum_{i \in \mathcal{I}} J^{L}_i(\bm{x}_i)  \notag\\
s.t. \label{P6} \quad  &
 \sum_{i=0}^I \bm{A}^c_i \bm{x}_i= \bm{b}^c. \tag{$\mathcal{P}^{''}_L$}\\ 
& \label{eq:18c} \bm{x}_i \in \mathcal{\bm{X}}_i, \forall i \in \mathcal{I} \cup\{0\}.\notag
\end{align}
where $\bm{x}_i\!=\![(\bm{x}_i(k))^T]_{k \in \mathcal{H}}^T$ , $\bm{x}_i(k)\!\!=\!\!\big(C_i(k), m^{z}_i(k), Z_i(k)\big)^T$ concatenating  the decision variables of zone $i$. 
 $J^{L}_i(\bm{x}_i)=\sum_{k \in \mathcal{H}}  \big(m^{z}_i(k)-m^{z, U}_i(k)\big)^2$ represents the local objective function of zone $i$. 
 $\sum_{i=0}^I \bm{A}^c_i \bm{x}_i= \bm{b}^c$  accounts for  the coupled linear constraints \eqref{total air flow rate bound} with an additional slack variables $\bm{x}_0(k) \geq 0$ introduced at each stage (transform the inequality constraints to equality constraints). 
$\mathcal{\bm{X}}_i$ ($\forall i \in \mathcal{I}$)  indicates the local constraints (\ref{CO2 ranges}),  \eqref{eq:11}, (\ref{13a})-(\ref{13e}) corresponding to zone $i$ and particularly $\mathcal{X}_0=\{\bm{x}_0 | \bm{x}_0 \geq \bm{0}\}$.  We have 
$\bm{A}^c_i= \small \left( \begin{array}{ccccccc}   
0&1& 0 &0 & 0& 0  &\cdots  \\  
0 &   0    &0   &0 &1 &0 &\cdots  \\
\cdots  & \cdots & \cdots &\cdots  & \cdots & \cdots  & \cdots  \\
\end{array} \small\right) \in \mathbb{R}^{H \times 3H}$ ($\forall i\in\mathcal{I}$),  
 and $\bm{b}^c=\big({m}^{\max}, {m}^{\max}, \cdots, {m}^{\max}\big)^T \in \mathbb{R}^H$. 

Following the standard procedure of ADAL \cite{chatzipanagiotis2015augmented}, we  have the augmented Lagrangian function: 
{\small 
\begin{eqnarray}
	\begin{split}
	\mathbb{L}_{\rho}\big( \bm{x}_0, \!\bm{x}_1, \!\cdots, \bm{x}_I, \bm{\alpha}\big)\!=&\! \!\sum_{i \in \mathcal{I}}J^{L}_i \!+\!\bm{\alpha}^T \big(\sum_{i=0}^I \bm{A}^c_i \bm{x}_i\!-\! \bm{b}^c\big)\\
	&\!+\!\frac{\rho}{2} \Vert \sum_{i=0}^I \bm{A}^c_i \bm{x}^i\!-\bm{b}^c \Vert^2\\
	\end{split}
\end{eqnarray} }
where $\bm{\alpha}=\big(\alpha_0, {\alpha}_1, \cdots, {\alpha}_{H-1}\big)^T$ are Lagrangian multipliers. $\rho  (\rho>0)$ is penalty parameter. 

Therefore,  we have the following primal problem with given Lagrangian multipliers $\bm{\alpha}$:
\begin{equation} \label{primal problem}
\begin{split}
      \min_{\bm{x}_0. \bm{x}^i,  \forall i \in \mathcal{I}} &\mathbb{L}_{\rho}\big(\bm{x}_0, \bm{x}_1, \cdots, \bm{x}_I,  \bm{\alpha}\big) \\
    s. t. 
        & \quad   \bm{x}_i \in  \mathcal{X}_i, ~~\forall i \in \mathcal{I}. \\
        &\quad  \bm{x}_0 \geq \bm{0}. 
\end{split}
\end{equation}

The main procedures of using ADAL to solve problem \eqref{P6} generally contain three steps: \emph{i)} solving the primal problem \eqref{primal problem}, 
\emph{ii)} updating the Lagrangian multipliers $\bm{\alpha}$, and \emph{iii)} updating the penalty factor $\rho$.   While the last two procedures are standard, we illustrate how to solve the primal problem \eqref{P6} in a distributed fashion.
Specifically,  we define $I+1$ agents, where Agent $1 \sim I$ correspond to the  $I$ zones and Agent  $0$ is a virtual agent for managing the slack decision variable $\bm{x}_0$. At each iteration $q$, we have the local objective functions  the agents:
 
{\small{
\begin{equation*}
\begin{split}
\mathbb{L}_{\rho}^0(\bm{x}_0, &\bm{x}^{q}_{-0}, \bm{\alpha})=\bm{\alpha}^T \bm{A}^c_0 \bm{x}_0+\frac{\rho}{2}\Vert  \bm{A}^c_0 \bm{x}_0+\sum_{i=1}^I \bm{A}^c_i \bm{x}_i^{ q} -\bm{b}^c \Vert^2\\
\mathbb{L}_{\rho}^i(\bm{x}_i, &\bm{x}_{-i}^{q}, \bm{\alpha})=J^{L}_i+ \bm{\alpha}^T \bm{A}^c_i \bm{x}_i+\frac{\rho}{2} \Vert  \bm{A}^c_i\bm{x}_i+\sum_{\mathclap{j\in \mathcal{I} \cup\{0\}, j\neq i}} \bm{A}^c_j \bm{x}_j^q -\bm{b}^c \Vert^2,\\
& \quad \quad \quad \quad  \forall  i \in \mathcal{I}. \\     
\end{split}
\end{equation*}
}}

It' clear that  we can obtain the estimated zone mass flow rates $m^z_i(k)$ and  zone CO$_2$ $C_i(k)$ by solving problem \eqref{P6}. Afterward, we invoke the procedure to update the estimation of $C_z(k)$  according to  \eqref{CO2 dynamics of supply air}.

Overall,  we present the details to solve  problem \eqref{P4} based on ADAL  in 
 \textbf{Algorithm} \ref{ADMM}.  We use the superscripts $p$ and $q$ to denote the iteration corresponding to solving the relaxed problem \eqref{P6} and updating the estimation of $C_z(k)$, respectively. 
For notation,  $\bm{x}^q=[ (\bm{x}_j^{q})^T]^T_{j \in \mathcal{I} \cup \{0\}}$ represent the augmented control and state trajectories for all agents while $\bm{x}_{-i}^{q}=[ (\bm{x}_j^{q})^T]^T_{j \in \mathcal{I} \{0\}, j \neq i}$  excluding Agent $i$.  
For the ADAL method, we  uses the residual error of coupled constraints: 
$r_q(\bm{x})= \Vert \sum_{i =0}^I \bm{A}^c_i \bm{x}_i^{q}\!-\!\bm{b}^c  \Vert \leq \epsilon^{\text{in}} $ (\textbf{C1}) 
 as  the stopping criterion. For the estimation of CO$_2$ concentration for the supply air, we evaluate  the  deviations  of any two successive estimation: 
${\big \Vert  \bm{C}^{p+1}_z-\bm{C}^{p}_{z}\big \Vert }\leq \epsilon^{\text{out}}$ (\textbf{C2}), 
where $\epsilon^{\text{in}}$ and $\epsilon^{\text{out}}$ are small positive thresholds.

\begin{algorithm}[h] 
	\caption{Solve problem \eqref{P6} based on ADAL } \label{ADMM}
	\begin{algorithmic}[1]
\State \textbf{Initialize} $ p \leftarrow 0$,  $\bm{C}^0_z$  and $\bm{d}_r$. 
	    \State \quad  \textbf{Initialize}   $ q\! \leftarrow\! 0$,  $\!\bm{\alpha}^0$, and  $\bm{x}^0_i$ ($\forall i \!\in \! \mathcal{I}\! \cup \! \{0\} $), $\bm{C}_z=\bm{C}^p_z$.    
		\State \quad  \textbf{for} {$i \in \mathcal{I} \cup\{0\}$}, \textbf{do}
		\begin{equation}
		\begin{split}
		&\bm{x}_i^{q+1}\!\!=\!\arg \min_{\bm{x}_i} \mathbb{L}^i_{\rho} (\bm{x}_i, \!\bm{x}_{-i}^{q},  \bm{\alpha}^q) ~~s. t.  ~ ~ \bm{x}_{i} \in \mathcal{X}_i.\\
		\end{split}
		\end{equation}		
		\State \quad  \textbf{end for}
		\State  \quad ~Update the Lagrangian multipliers:
		$$\bm{\alpha}^{q+1}\!=\!\bm{\alpha}^q\!+\!\rho\big( \sum_{i=0}^I \bm{A}^c_i \bm{x}_i^{q+1}\!-\!\bm{b}^c \Big) $$
	\State  \quad  If (\textbf{C1}) is satisfied, stop with  $\bm{x}^p = \bm{x}^{q+1}$, otherwise $q \rightarrow q+1$  \hspace*{3mm} and go to Step   3. 
		\State Estimate $\bm{C}^{p+1}_{z}$ according to
		{\small{
		 \begin{equation}
		 	\begin{split}
		 	\bm{C}^{p+1}_z=(1-\bm{d}_r)\bm{C}_o+\bm{d}_r \frac{\sum_{i \in \mathcal{I} }  \bm{m}^{z, p}_i  \bm{C}_i^{p}}{\sum_{i \in \mathcal{I}}\bm{m}^{z, p}_i}. \\
		 	\end{split}
		 \end{equation}
		 }}
		 \State If (\textbf{C2}) is satisfied, stop with $\bm{x}^p$, otherwise set $p \rightarrow p+1$ and go to Step 2.  
		 \Ensure {\small{$\bm{x}^{*}\!\!=\!\![(\bm{x}^p_i(k))^T]_{k \in \mathcal{H}}^T$ , $\bm{x}^p_i(k)\!\!=\!\!\big(C^p_i(k), m^{z, p}_i(k), Z^p_i(k)\big)^T $. }}
	\end{algorithmic}
\end{algorithm}

\emph{Recursive feasibility}: 
Recall that  the bilinear terms $Z_i(k)=m^{z}_i(k) C_i(k)$  are relaxed  in problem \eqref{P5} (or \eqref{P6} ), therefore the recursive feasibility remains to be addressed. 
We propose a heuristic method (see \textbf{Algorithm} \ref{Heuristic method}) to recover a 
 control input  $\hat{\bm{x}}=[ (\hat{\bm{x}}_i(k))^T]_{k \in \mathcal{H}}^T$ 
 from  the optimal solution of problem \eqref{P6}, which can ensure the recursive feasibility and  \emph{near} energy saving performance. 
The main idea is to preserve the computed zone mass flow rates: $\hat{m}^z_i(k)=m^{z, *}_i(k)$ ($\forall i\in\mathcal{I}, k \in \mathcal{H}$)  as they determine the HVAC's energy cost.    In such setting,  
the desirable CO$_2$  at each executed epoch $t$ will be sustained.  As the method is  deployed in MPC framework, the desirable CO$_2$ over the day (optimization horizon $\mathcal{T}$) will be achieved along the computing epoch. 
We illustrate the \emph{feasibility issue} by induction. For notation, we indicate the real zone zone CO$_2$ trajectories as 
 $[C_{i, t}]_{t \in \mathcal{T}}$ ($\forall i \in {\cal I}$),  which suppose to be obtained at the end of optimization horizon $\mathcal{T}$.  At each execution instant $t$, we can obtain the current zone CO$_2$ measurements  $C_i(0) = C_{i, t}$ ($\forall i \in \mathcal{I}$).   Suppose at time $t$ we have $C_{i, t} \leq C^{\max}_i$ ($\forall  i \in \mathcal{I}$),  it suffices to address the \emph{feasibility issue}  by  proving $C_{i, t+1} \leq C^{\max}_i$ ($\forall i \in \mathcal{I}$) as below.
\begin{equation*}
\begin{split}
C_{i, t+1}&=   C_i(1) \\
&=C_i(0)+E_{i}(0) \hat{m}^{z}_i(0) +F_i(0) \hat{m}^{z}_i(0) C_i(0)+G_i(0)\\
& = C_i(0)+E_{i}(0) m^{z, *}_i(0) +F_i(0) m^{z, *}_i(0) C_i(0)+G_i(0)\\
& = C_i(0)+E_{i}(0) m^{z, *}_i(0) +F_i(0) Z^{*}_i(0) +G_i(0)\\
& \leq C^{\max}_i, \forall i \in \mathcal{I}. 
\end{split}
\end{equation*}
where the first two equalities  are  directly from zone CO$_2$ dynamics  \eqref{CO2 dynamics2}. The third  and fourth equality are  derived from $\hat{m}^z_i(k) = m^{z, *}_i(k)$ and  \eqref{13a}.  The inequality is deduced from constraint \eqref{CO2 ranges}. 


\begin{algorithm}[h] 
	\caption{Recover Recursive Feasibility} \label{Heuristic method}
	\begin{algorithmic}[1]
		\Require {\small{$\bm{x}^{*}\!\!=\!\![(\bm{x}^p_i(k))^T]_{k \in \mathcal{H}}^T$ , $\bm{x}^p_i(k)\!\!=\!\!\big(C^p_i(k), m^{z, p}_i(k), Z^p_i(k)\big)^T $ }}  (from \textbf{Algorithm} \ref{ADMM}).
		\State Set  $\hat{C}_i(0)=C_i(0)$ and  $\hat{T}_i(0)=T_i(0)$ ($\forall i\in\mathcal{I}$). 
		\For{$k \in \mathcal{H}$}
		\For{$i\in \mathcal{I}$}
		\State Set $ \hat{m}^{z}_i(k)=m^{z, *}_i(k)$ and  $\hat{Z}_i(k)=\hat{m}^{z}_i(k) \hat{C}_i(k)$.
		\State Estimate  $\hat{C}_i(k+1)$ and $\hat{T}_i(k+1)$ by
		{\small { 
				\begin{displaymath}
				\begin{split}
				&\quad \quad \hat{C}_i(k+1)=\hat{C}_i(k)+\hat{E}_i(k) \hat{m}^{z}_i(k) +F_i(k) \hat{Z}_i(k)+G_i(k), \\
				&\quad \quad \hat{T}_i(k+1)=A_{ii}\hat{T}_i(k)\!+\!\!\sum_{j\in \mathcal{N}_i} A_{ij}\hat{T}_j(k) \\
				&\quad \quad \quad \quad \quad \quad \quad \quad + C_{ii} \hat{m}^{z}_i(k) (\hat{T}_i(k)-T_c)  +\!D_{i}(k),  \\
				& \quad \quad \textrm{with}~ \hat{E}^{z}_i(k)={\hat{C}_z(k) \Delta_t}/{m_i}, \textrm{and}~  \hat{C}^z(k)=(1-d^*_{r}(k)) C_o(k)\\
				& \quad \quad+d^{*}_{r}(k) \frac{\sum_{i \in \mathcal{I}} \hat{m}^{z}_i(k) \hat{C}_i(k)}{\sum_{i \in \mathcal{I}} \hat{m}^{z}_i(k)}, \forall i \in \mathcal{I}, k \in \mathcal{H}. \\
				\end{split}
				\end{displaymath}
		}}
		\EndFor
		\EndFor
		\Ensure  $\bm{\hat{m}}^{z}_i\!=\![\hat{m}^z_i(k)]_{k \in \mathcal{H}}$, $\bm{\hat{C}}_i=[\hat{C}_i(k)]_{k \in \mathcal{H}}$,  and $\bm{\hat{T}}_i=[\hat{T}_i(k)]_{k \in \mathcal{H}}$. 
	\end{algorithmic}
\end{algorithm}

\emph{Second phase of LLC}: As discussed previously,  it may occur that the lower comfortable zone temperature bounds may be  violated  for solving problem \eqref{P4}. In such situation,  we are required to invoke the   second \emph{phase} of LLC  to increase  ventilation rate  (decrease $d_r$). In principle, the regulating  step-size should be  determined by the temperature violation amplitude,   however it's difficult to coordinate such two different dimensions and we thus use a  small constant step-size $\Delta d_r$.  The overall framework of the proposed 
TLDM which embeds the ULC  and LLC is presented in \textbf{Algorithm} \ref{hierarchical decentralized method} . We use $l$  to denote  the iteration.  Function $\mathbb{I} (A)$ is an indicator function, where we have $\mathbb{I} (A)=1$  with true condition $A$, otherwise $\mathbb{I}(A)=0$.

\begin{algorithm}[h] 
	\caption{Two-level Distributed Method (TLDM)} \label{hierarchical decentralized method}
	\begin{algorithmic}[1]
					\State \textbf{Initialize} $ l \leftarrow 0$ and  $\bm{d}^0_r$.
		\State  $[\bm{m}^{z, L}_i]_{i \in \mathcal{I}}=ULC(\bm{d}^l_r)$ \cite{yu2018TASE}.
	    \State $[\bm{\hat{m}}^{z}_i, \bm{\hat{C}}_i, \bm{\hat{T}}_i]_{i \in \mathcal{I}}=LLC( [\bm{m}^{z, L}_i]_{i \in \mathcal{I}}, \bm{d}^l_r)$(\textbf{Algorithm} \ref{ADMM}-\ref{Heuristic method}). 		
\State  If $\bm{\hat{T}}^i\geq {T}^{\min}_i$, then stop, otherwise continue.
		\State Update ventilation rate  $\bm{d}_r$:
		{\small{ 
		\begin{displaymath}
		\bm{d}_r^{l+1}(k)=\bm{d}^l_{r}(k)-\Delta d_{r}  \mathbb{I} (\bm{\hat{T}}_{i}(k+1) <{T}^{\min}_i )  , ~\forall k \in \mathcal{H}. 
		\end{displaymath}
		}}
			\State Set $l\rightarrow l+1$ and go to \textbf{Step} 2.
	\end{algorithmic}
\end{algorithm}

\vspace{-5mm}
\section{Application}
We evaluate the performance of the TLDM on multi-zone commercial HVAC system via  simulations.  We first study  the energy saving performance  and computational advantage of the method via  a \emph{benchmark} ($5$ zones).
  After that  the capability and scalability of the method to \emph{medium} scale (10,20 zones) and  \emph{large} scale (50,100 zones)  are illustrated.


\subsection{Benchmark}
\begin{figure}[h]
	\centering 
	\includegraphics[height= 3.2cm,width=2.5 in]{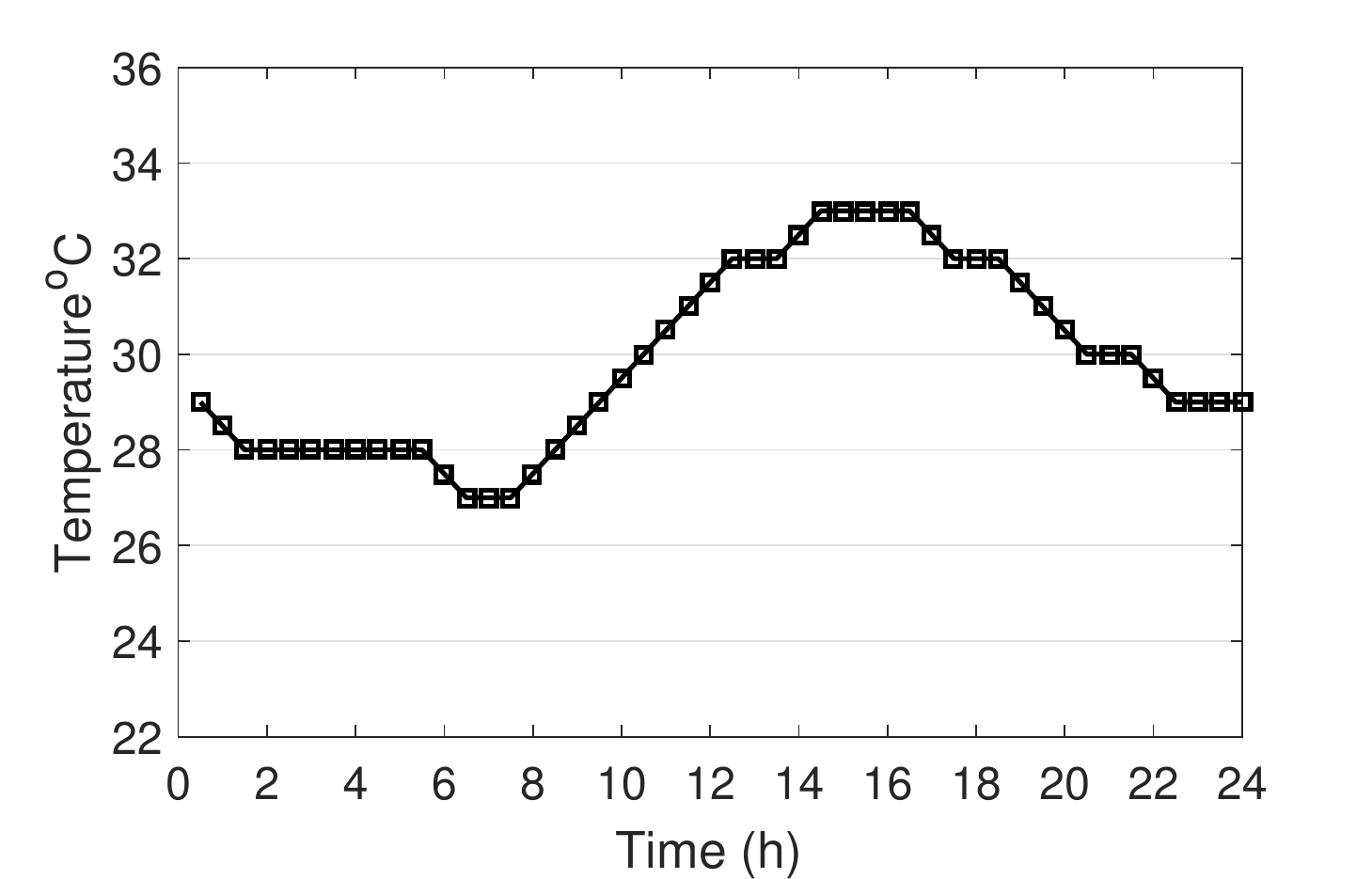}
	\includegraphics[height=3.2cm,width=2.5 in]{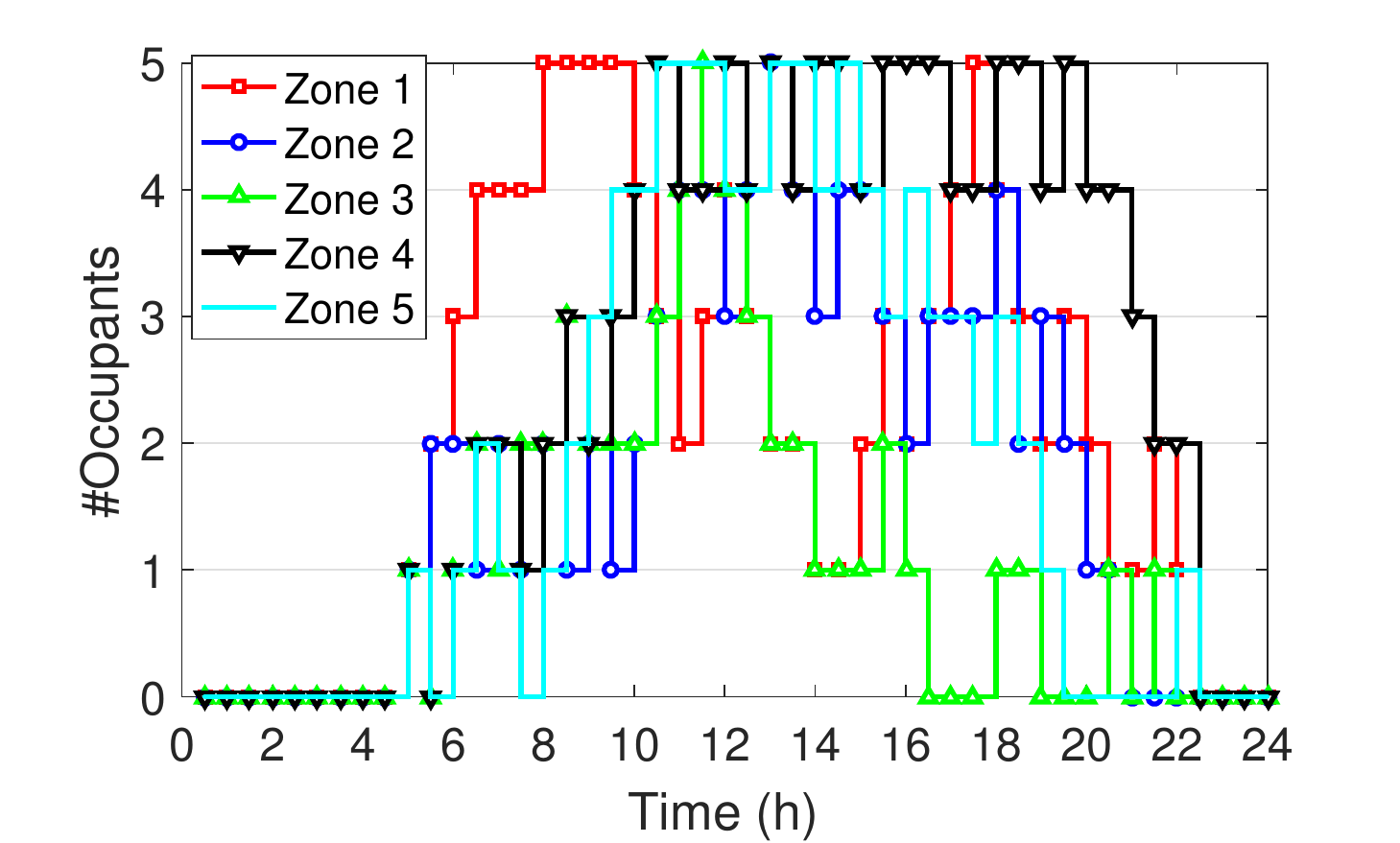}
	\caption{(a) Outdoor temperature.   (b) Zone occupancy. } \label{OTOC}
\end{figure}

This part considers a $5$-zone \emph{benchmark}.  We select the  general comfortable zone temperature range  $[24, 26]^\circ$C  and zone CO$_2$ range $[0, 800]$~ppm. 
The set-point temperature of AHU is  $T_c=15^\circ$C. 
We assume the zones are spatially next to each other $1 \leftrightarrow 2 \leftrightarrow 3 \leftrightarrow 4 \leftrightarrow 5 \leftrightarrow 1$.
The initial zone temperature is  set as $[29, 30, 31, 30, 29]^{\circ}$C (zone 1-5). 
The predicted outdoor temperature and zone occupancy are shown in  Fig.~\ref{OTOC}.
The HVAC's energy cost  is calculated according to the time-of-use (TOU) price in Singapore \cite{xu2017pmv}. The other parameters refer to TABLE \ref{System  Parameters}.
\begin{table}[h]
	\setlength{\abovecaptionskip}{-4pt}
	\setlength{\belowcaptionskip}{4pt}
	\scriptsize
	\centering
	\caption{Simulation Parameters} \label{System  Parameters}
	\begin{tabular}{p{2cm}p{2cm}p{2cm}}
		\toprule[1.0pt]
		Param.  & Value   &Units\\
		\hline
		$C_i (i\in \mathcal{I})$         &  $1.5 \times 10^3$      & \si{\kilo\joule\per\kelvin}               \\
		$c_p$                                   & $1.012$                       & \si{\kilo \joule \per \kilogram- \kelvin}\\    
		$R_{oi}$                                & $50$                             & \si{\kilo\watt\per\kelvin}            \\
		$R_{ij} (i, j \in \mathcal{I})$   & $14$                             &\si{\kilo\watt\per\kelvin}             \\
		$\kappa_f$                                    & $0.08$                          &-                          \\       
		$\eta$                                   & $1$                                &-                         \\
		$C_g$                                   & $40$                             & \si{\gram\per\hour}             \\
		$\Delta d_r$                           & $0.05$                            &-                         \\
		${m}^{z, \min}_i$               & $0$                                 & \si{\kilogram\per\hour}             \\
		${m}^{z, \max}_i$                  &$0.5$                              &  \si{\kilogram\per\hour}            \\
		\bottomrule[1.0pt]
	\end{tabular}
\end{table}

 In the \emph{benchmark}, we compare the proposed method  with \emph{i)} distributed Token-based scheduling strategy (DTBSS) \cite{radhakrishnan2016token},  {\em{ii)}}  centralized method,  
 {\em{iii)}} the commonly-used DCV strategies  \cite{sun2011situ, shan2012development} and {\em{iv)}} sequential quadratic programming (SQP) \citep{nocedal2006sequential}.   
Akin to most existing works, the DTBSS lacks IAQ management \cite{radhakrishnan2016token} and we therefore fix the ventilation rate in DTBSS  as $d_r(k)=d_r^{\max}$ ($\bm{k} \in \mathcal{H}$) to achieve the energy cost saving target.  For the centralized method,  we obtain the optimal solution  by solving the non-linear optimization problem (\ref{obj}) using the IPOPT solver  embedded in MATLAB \cite{kawajir2006introduction}. 
The DCVs calculate  the amount of fresh air by the zone occupancy and zone area \cite{sun2011situ, shan2012development}, i.e.,
 \begin{equation} \label{DCV}
 	\begin{split}
 	m^{z, \textrm{fresh}}_i(k)=N_i(k) R_p+A_i  R_a, ~\forall k \in \mathcal{H}.
 	\end{split}
 \end{equation}
 where $A_i$ denotes the area of zone $i$. $R_p$  and $R_a$ denote the average occupancy and space ventilation rate. 
This paper considers two DCVs: \emph{i)} DCV I: calculating  zone fresh air flow rates  based on zone occupancy ($R_p \geq 0$,   $R_a=0$),  and \emph{ii)} DCV II: 
computing zone fresh air flow rates by zone  occupancy and  space ($R_p \geq 0$,   $R_a \geq0$). 
 
 For single-zone case, the ventilation rate  ($d_r$) of  DCVs can be straightforwardly determined by the fresh air infusion and mass flow rate. Nevertheless, for multi-zone case,  the calculation of  ventilation rate needs to account for zone diversity \cite{shan2012development}: 
 {\small{
 \begin{eqnarray} \label{CO2-based DCV}
 	\begin{split}
 	&m^{z, \textrm{fresh}}(k) =\sum_{i \in \mathcal{I}} m_i^{z, \textrm{fresh}}(k) ,  ~ m^z(k)=\sum_{i \in \mathcal{I}}  m^{z}_i(k), \\
 	&Z(k)=\max_{i \in \mathcal{I}} \Big\{ \frac{m^{z, \textrm{fresh}}_i(k)}{m^{z}_i(k) }\Big\},  \quad X(k)=\frac{m^{z, \textrm{fresh}}(k)}{m^z(k)},\\ 	
 	&Y(k)=\frac{X(k)}{1+X(k)-Z(k)}, \quad  d_{r}(k)=1-Y(k), ~\forall k \in \mathcal{H}.\\
 	\end{split}
 \end{eqnarray}
 }}
 
One critical issue to be noticed is that  the DCVs only account for IAQ by determining the amounts of fresh air infusion, the thermal comfort which corresponds to the zone mass flow rates is not considered.    Therefore,  we use the ULC of  TLDM to calculate  zone mass flow rates $m^{z}_i(k)$  in DCVs for fair comparisons. 
Therefore, similar to our TLDM, the DCVs correspond to two alternative procedures: obtain the zone mass flow rates by solving problem  (\ref{P2})  (start with $d_r(k)=d^{\max}_r$) and  amending the ventilation rate $d_{r}(k)$ according to (\ref{CO2-based DCV}).


  \begin{table}[h]
 	\setlength{\abovecaptionskip}{-4pt}
 	\setlength{\belowcaptionskip}{4pt}
 	\scriptsize
 	\centering
 	\caption{Performance comparisons} \label{performance comparison}
 	\begin{tabular}{ p{1.3cm}p{0.5 cm} p{0.5 cm}p{1.0cm}p{0.7cm}p{0.4cm}p{0.6cm}}
 	\toprule[1.0pt]
 	 \multirow{2}{*}{Method}                        &$R_p $    &  $R_a $    & Cost   &Time    &      \multirow{2}{*}{TC} & \multirow{2}{*}{IAQ}\\
 	&  &{\scriptsize (L/p)}          & {\scriptsize (\si{\liter\per\meter^2}) }    &(s)        &  &\\
 	\hline  
 	DTBSS  &  -                     & -                             &    $245.03$   &  $2.11$    & \textbf{Y}& \textbf{N}  \\
 	Centralized            &  -                     &-                              &     $247.15$  & $575.07$  & \textbf{Y}  & \textbf{Y}  \\
 	SQP                      & -                      &-                              &         $275.91$     &  $151.26$       &             \textbf{Y}       &      \textbf{Y}     \\
 	DCV  I                        & $21$              &  $0$                       &    $276.23$  &  -  & \textbf{Y} & \textbf{Y}   \\
 	DCV II                         &  $16$             & $0.04$                   &  $274.99$  &  -  & \textbf{Y}   & \textbf{Y}   \\
        TLDM    &-                       &-                               &   $257.02$  & $5.21$  & \textbf{Y}  & \textbf{Y}  \\
 	\bottomrule[1.0pt]
 	\end{tabular}
 	\begin{tabular}{p{2cm}p{5cm}}
 	\textbf{N}=No, \textbf{Y}=Yes. & \\
 	\end{tabular}
 \end{table}

\begin{figure}[h]
	\setlength{\abovecaptionskip}{-4pt}
	\setlength{\belowcaptionskip}{4pt}
	\centering
	\includegraphics[width=3.6 in]{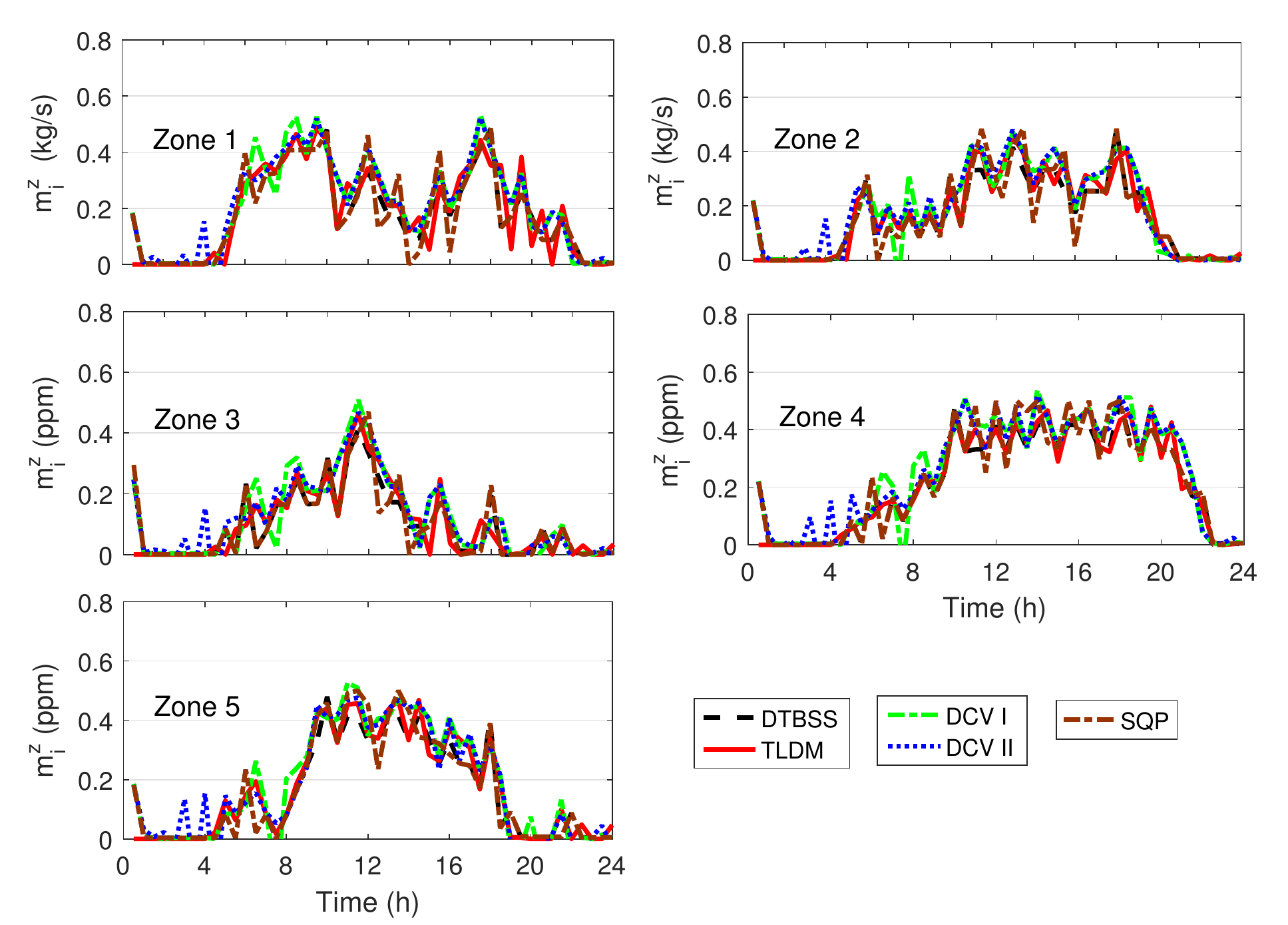}\\
	\caption{Zone mass flow rate (\emph{Benchmark}). } \label{zone air flow rate}
\end{figure} 

\begin{figure}[h]
	\setlength{\abovecaptionskip}{-4pt}
	\setlength{\belowcaptionskip}{4pt}
	\centering
	\includegraphics[width=3.4 in]{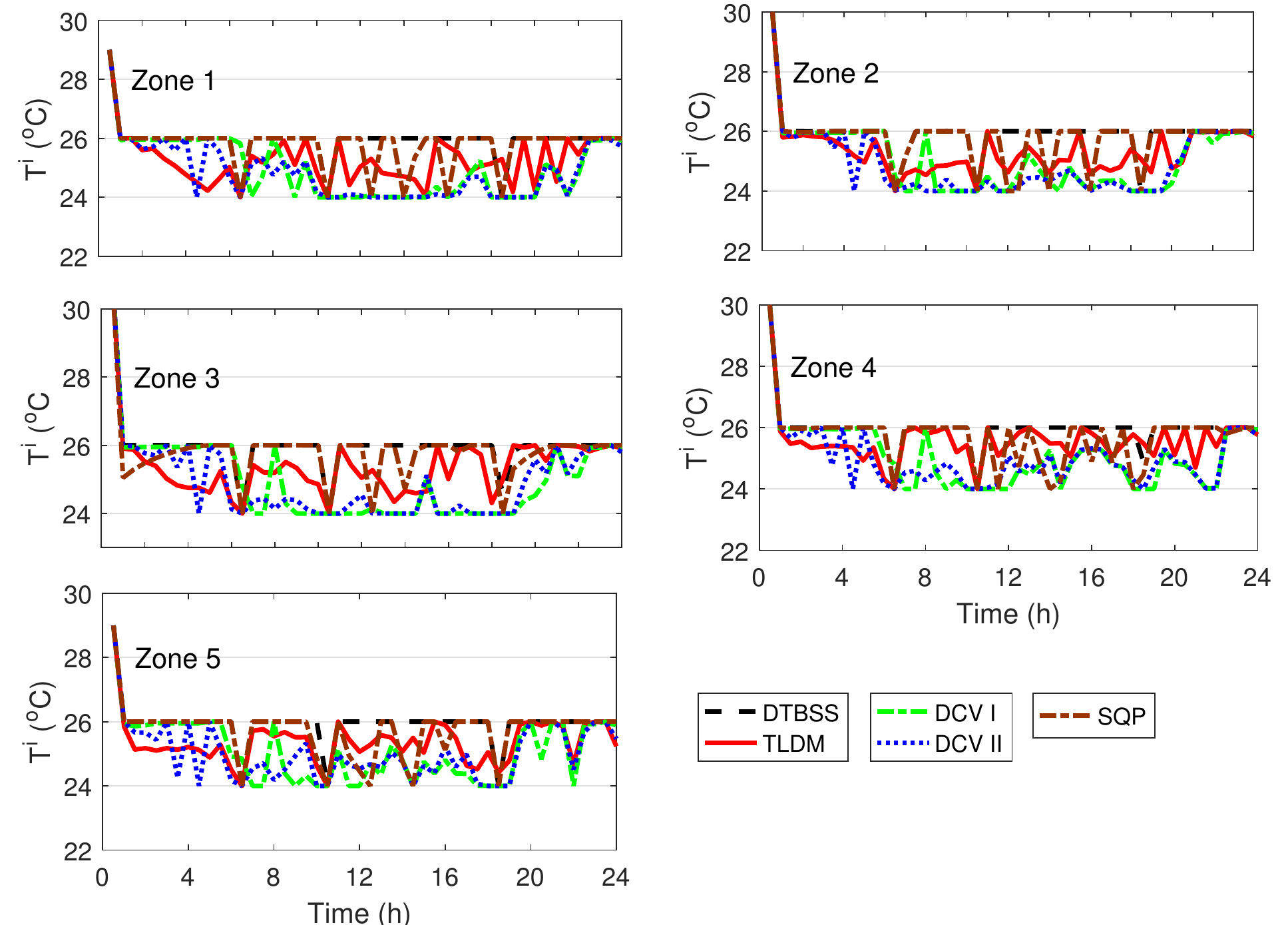}\\
	\caption{Zone temperature (\emph{Benchmark}). } \label{zone temperature}
\end{figure} 

\begin{figure}[h]
	\setlength{\abovecaptionskip}{-4pt}
	\setlength{\belowcaptionskip}{4pt}
	\centering
	\includegraphics[width=3.4 in]{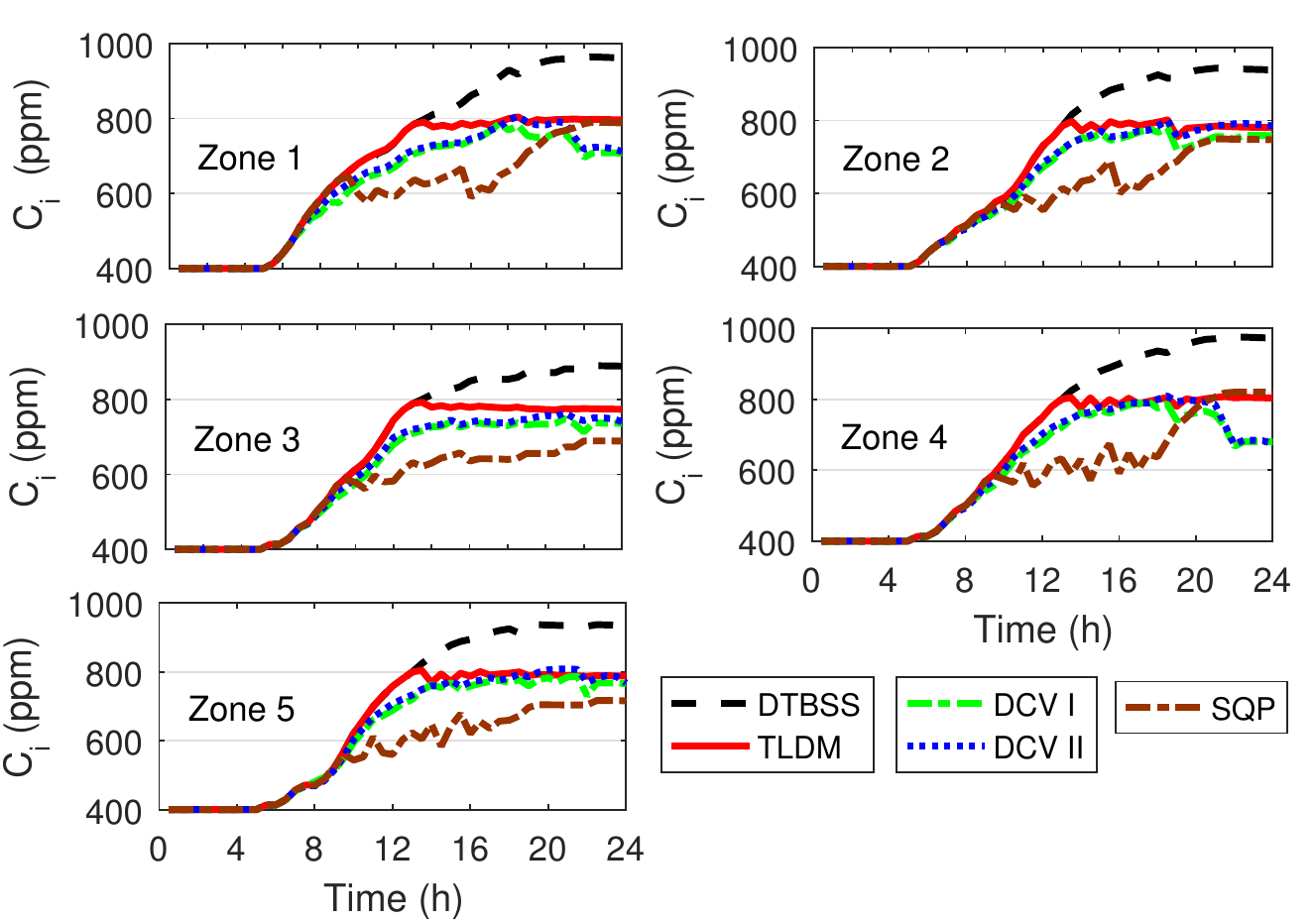}\\
	\caption{Zone CO$_2$ (\emph{Benchmark}). } \label{zone CO2}
\end{figure}

\begin{figure}[h]
	\setlength{\abovecaptionskip}{-4pt}
	\setlength{\belowcaptionskip}{4pt}
	\centering
	\includegraphics[width=3.2 in, height = 1.8 in]{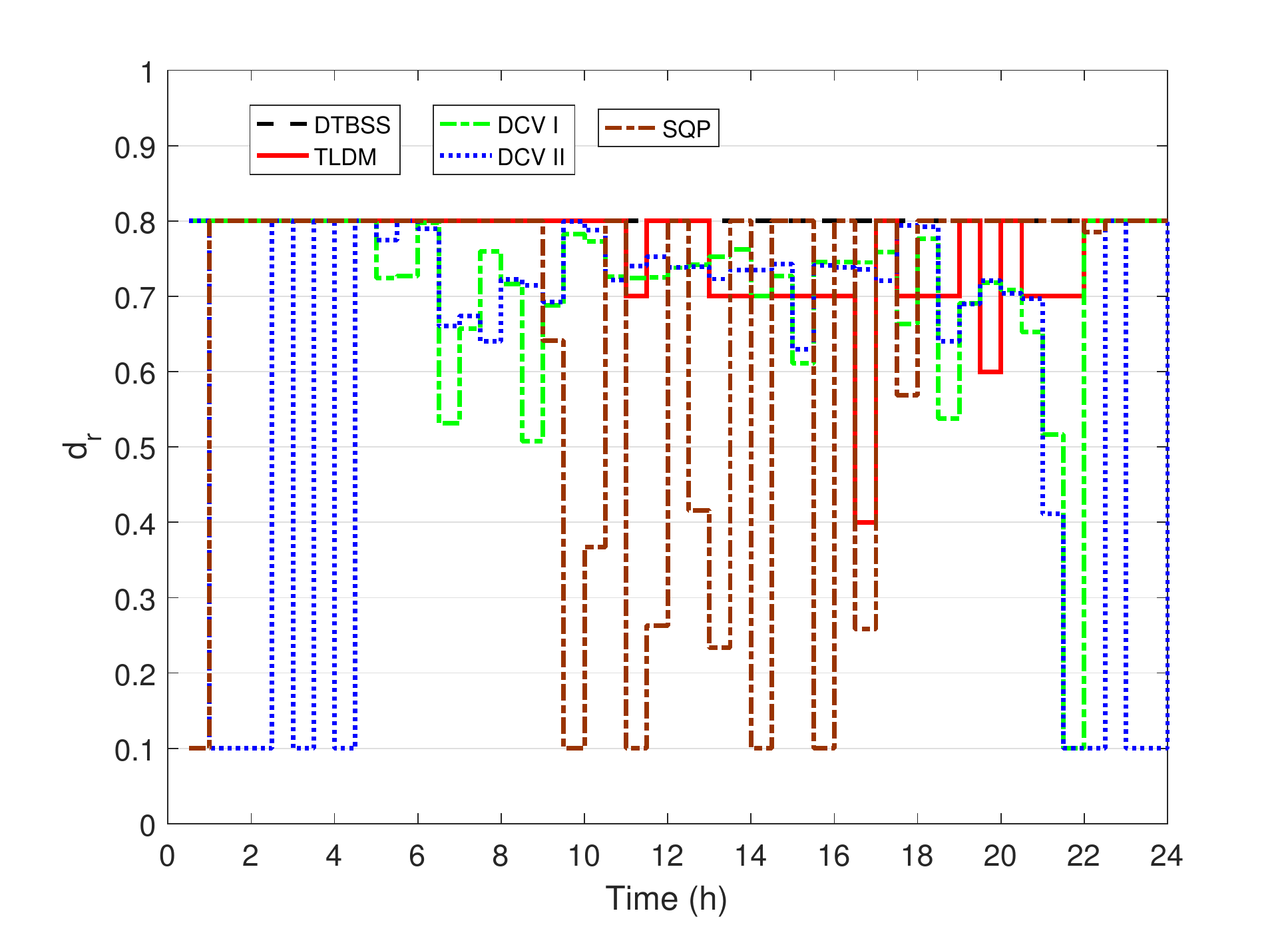}\\
	\caption{Ventilation rate $\bm{d}_{r}$ (\emph{benchmark}).} \label{dr}
\end{figure}

We present  the  energy cost,  average computation time of each computing epoch {\footnote{MATLAB R2016a on PC  with Intel(R) Core(TM) i7-5500U CPU @2.40GHZ  processor.},  satisfaction of thermal comfort (TC) and  IAQ   induced by  the different methods  in TABLE  \ref{performance comparison}.  
In particular,  we exclude the DCVs while studying the computation time as they depend on \emph{off-line} regulation for the occupancy and space ventilation rates, which is time-consuming. 
We display the zone mass flow rates,  zone temperature, zone CO$_2$  and  ventilation rate  in Fig. \ref{zone air flow rate}-\ref{dr}.
First of all, we find the zone thermal comfort is ensured by each method  (Fig. \ref{zone temperature}) but not necessary  for IAQ (Fig. \ref{zone CO2}) as  we observe the desirable zone CO$_2$ are out of range $[0, 800]$~ppm for  DTBSS. This is not surprising as  the DTBSS lacks  IAQ management. 
Therefore,  the method  favors  low ventilation rate (see Fig. \ref{dr})  to save energy cost and fails to ensure IAQ. 
Further, we study the energy cost  and computational efficiency of different methods.  From TABLE \ref{performance comparison}, we find  the TLDM 
provides about  $7.0\%$ lower  energy cost  compared with the DCVs and the SQP.   
Since we observe close peaks over the zone temperature  (Fig. \ref{zone temperature}) and zone CO$_2$ (Fig. \ref{zone CO2}) trajectories  for these methods, we conclude that the TLDM can maintain the same thermal comfort and IAQ but with less energy cost.  
As for the computational efficiency,  the average computation time of TLDM  for each executive epoch is about 5.21\si{\second} (in parallel) with a slight increase over  the DTBSS (2.11\si{\second}). This is mainly  attributed to the LLC to achieve IAQ. However,  the TLDM  obviously outperforms SQP in computation efficiency. 
While compared with centralized method, we imply the sub-optimality of TLDM in energy cost  is around $4\%$ in the \emph{benchmark}.  Nevertheless, the  computational benefit of TLDM is  significant as the average computation time is reduced  from 575.07s to 5.21s.


\subsection{Scalability}
In this part, the proposed TLDM is  applied  to  \emph{medium} (10,20 zones) and \emph{large} (50,100 zones) scale cases.  Considering that centralized method  and  SQP  are now computationally intractable, we compare the TLDM  with the other three methods (i.e.,  DTBSS, and DCV I, II).  For each case, we randomly generate a network to represent the spatial connectivity of the zones (the maximum number of adjacent zones zone is  set as  $4$).  In particular,  the space and occupancy ventilation rates for  the DCVs to achieve  the IAQ are documented in TABLE \ref{ventilation rates}, which are obtained from  \emph{off-line} regulation. 
The other parameters refer to the \emph{benchmark} in Section IV-A.

\begin{table}[h]
	\setlength{\abovecaptionskip}{-2pt}
	\setlength{\belowcaptionskip}{2pt}
	\scriptsize
	\centering
	\caption{Occupancy and Space Ventilation Rates in DCV I, II } \label{ventilation rates}
	\begin{tabular}{p{0.8 cm} | p{1.2cm}| p{1.2cm} p{1.4 cm} }
		\toprule[1.0pt]
       \multirow{2}{*} {\#zones}  & DCV I   &  \multicolumn{2}{c}{DCV II}\\
       \cline{2-4}
                                                 & $R_p$ (\si{\liter}/p)       &  $R_p$ (\si{\liter}/p)   & $R_a$ (\si{\liter\per\meter^2})\\
                                                 \cline{2-4}
          $10 $         & $19$          & $15$          & $0.03$       \\                     
          $20$          & $20$          & $19$          & $0.03$       \\
          $50$          & $21$          &   $19$        &  $0.03$      \\
          $100$        & $23$           &  $21$        & $0.03$       \\
			\bottomrule[1.0pt]
	\end{tabular}
\end{table}

Similarly, we investigate the energy cost and computational efficiency of different methods,  with the results  shown  in TABLE \ref{HVAC cost}. 
Compared with the DTBSS, we observe a minor  increase   in  energy cost  and computation time with the TLDM to achieve  IAQ. 
However,  the TLDM  is still computationally efficient and scalable by inspecting the average  computation time (e.g., $21.68$\si{\second} for $100$-zone case) \emph{versus} the decision epoch $30$\si{\minute}s.
Besides,  we observe similar gains of  the TLDM  over the DCVs in energy cost saving:  $8.0$-$9.8\%$ ({DCV I}) and $8.1$-$10.2\%$ ({DCV II}).  
Moreover,  except for the energy cost reduction, the TLDM is expected to  be more applicable over the DCVs as it doesn't depend on time-consuming \emph{off-line} regulation for  the  occupancy and  space ventilation rate.

\begin{table}[h]
	\setlength{\abovecaptionskip}{-2pt}
	\setlength{\belowcaptionskip}{2pt}
	\scriptsize
	\centering
	\caption{Performance comparisons} \label{HVAC cost}
	\begin{tabular}{p{0.8 cm} | p{0.6cm} p{0.6cm} p{0.5cm}|p{0.7cm} p{0.6cm} p{0.5cm}|    p{0.2cm} p{0.2cm}}
		\toprule[1.0pt]
		\multirow{3}{*}{Method}  & \multicolumn{6}{c|}{\emph{Medium}} & \multirow{3}{*}{TC} & \multirow{3}{*}{IAQ} \\
		\cline{2-7} 
		&    \multicolumn{3}{c|}{ $10$}   & \multicolumn{3}{c|}{$20$}   \\
		\cline{2-7}
		& Cost (s\$)   &  +/- ~~ (\%) & Time (s) & Cost (s\$)   & +/-~~ (\%)  & Time (s) \\
		\hline
		TLDM   & $407.32$     & -               &  $6.55$      &   $939.83$     &-                 & $8.03$     &  \textbf{Y}  & \textbf{Y} \\
		DTBSS &  $387.07$    &  -$4.97$    &  $2.35$     &  $887.12 $      & -$5.61$     &   $2.71$  &   \textbf{Y}  & \textbf{N} \\
		DCV I &    $447.42$   &  +$9.84$    &   -   &  $1015.50$       &+$8.05$    & -    & \textbf{Y}  & \textbf{Y}   \\
		DCV II &  $440.42$    & +$8.13$     &    -  &  $1020.60$      & +$8.59$     &   - &  \textbf{Y}  & \textbf{Y}\\
		\bottomrule[0.6pt]
	\end{tabular}
	
	\begin{tabular}{p{1cm}p{1cm}}
		& \\
	\end{tabular}
	
	\begin{tabular}{p{0.8 cm} | p{0.6cm} p{0.6cm} p{0.5cm}|p{0.7cm} p{0.6cm} p{0.5cm}|    p{0.2cm} p{0.2cm}}
		\toprule[0.6pt]
		\multirow{3}{*}{Method}  & \multicolumn{6}{c|}{\emph{Large}} & \multirow{3}{*}{TC} & \multirow{3}{*}{IAQ} \\
		\cline{2-7} 
		&    \multicolumn{3}{c|}{ $50$}   & \multicolumn{3}{c|}{$100$}   \\
		\cline{2-7}
		& Cost (s\$){\tiny{$\times 10^3$}}   &  +/- ~~ (\%)   & Time (s) & Cost (s\$){\tiny{$\times 10^3$}}  &   +/- ~~ (\%)  & Time (s) \\
		\hline
		TLDM   & $2.65$      & -                 & $13.28$ &    $5.80$    &-                   &  $21.68$  &  \textbf{Y}  & \textbf{Y} \\
		DTBSS &  $2.54$     &  -$4.15$     & $4.47$  &   $5.49$      &     -$5.34$    &    $6.90$ &   \textbf{Y}  & \textbf{N} \\
		DCV I &    $2.91$     &  +$9.81$   & -             &    $6.31$     & +$8.79$                 & -               & \textbf{Y}  & \textbf{Y}   \\
		DCV II &  $2.92$      & +$10.19$  & -             &  $6.36$       &    +$9.66$             &  -              &  \textbf{Y}  & \textbf{Y}\\
		\bottomrule[1.0pt]
	\end{tabular}
	
	\begin{tabular}{p{2cm}p{5cm}}
		\textbf{N}=No, \textbf{Y}=Yes. & \\
	\end{tabular}
\end{table}



\begin{figure}[h]
	\centering
	\setlength{\abovecaptionskip}{-4pt}
	\setlength{\belowcaptionskip}{4pt}
	\includegraphics[height=3.4 cm,width=3.0 in]{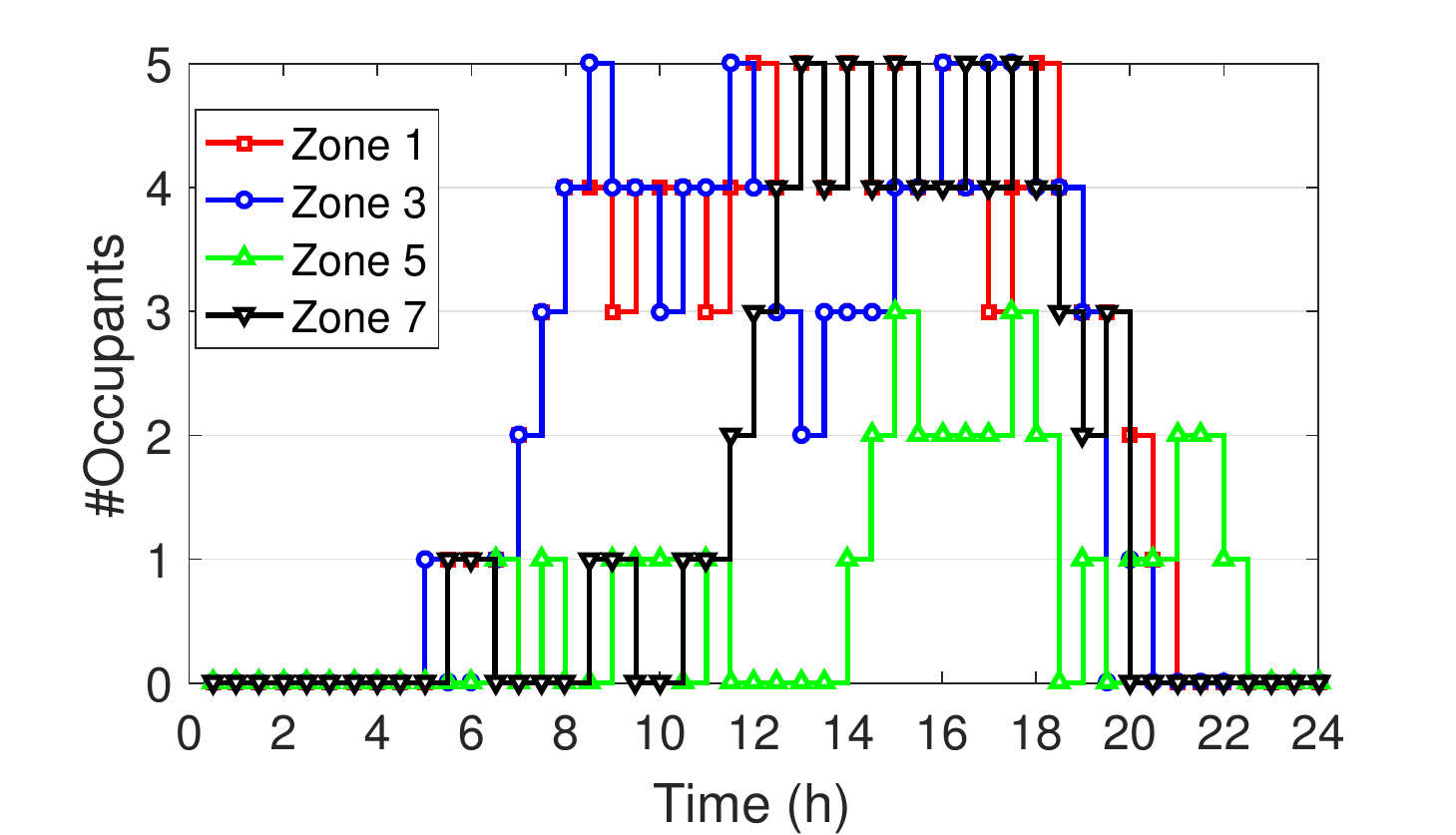}\\
	\caption{The dynamic occupancy for  the three  zones among $I=50$ zones.} \label{occupancy50}
\end{figure} 

\begin{figure}[h]
	\setlength{\abovecaptionskip}{-4pt}
	\setlength{\belowcaptionskip}{4pt}
	\centering
	\includegraphics[width=3.6 in]{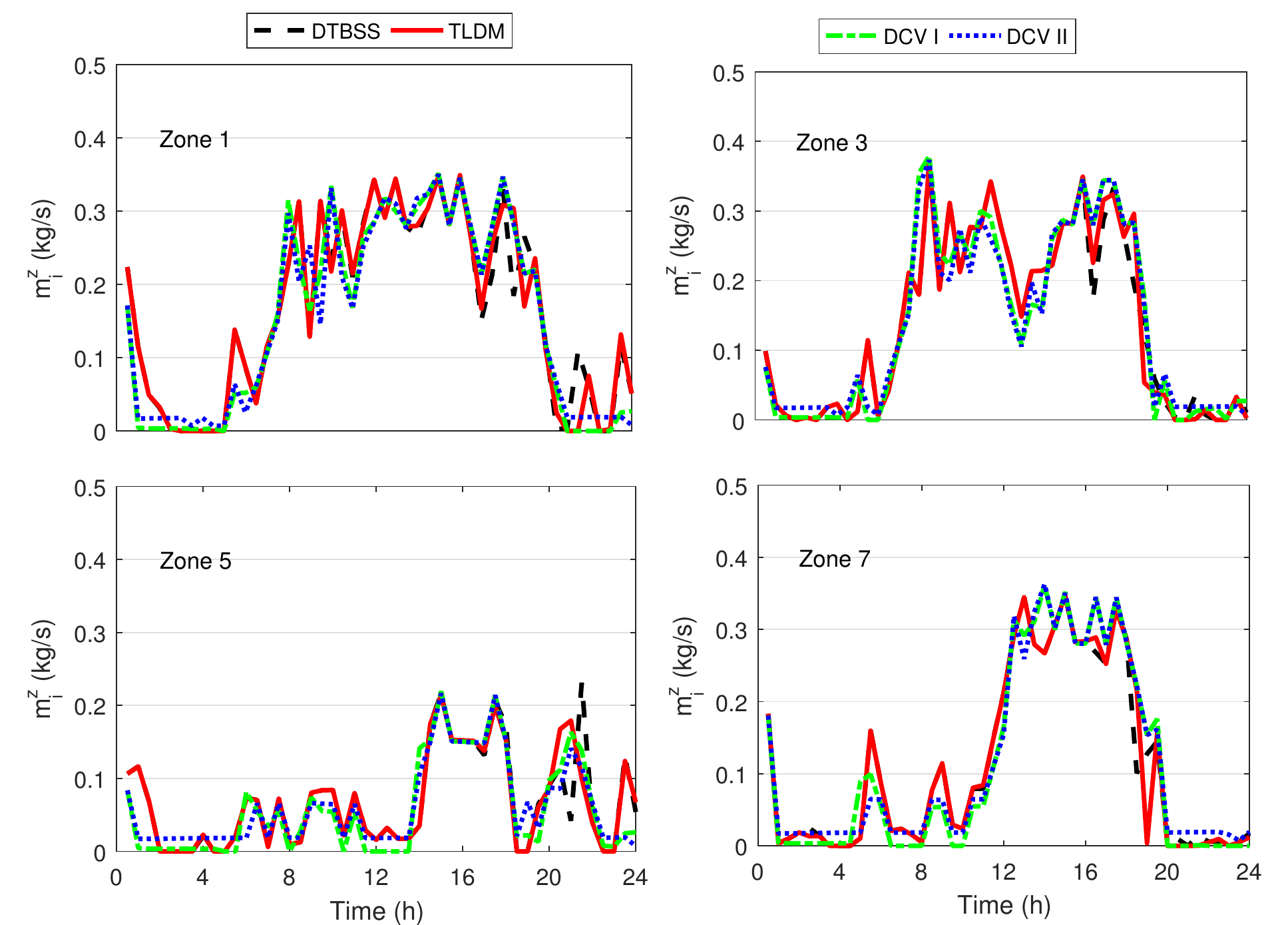}\\
	\caption{The zone air flow rates for the three zones among $I=50$ zones.} \label{zone air flow rate50}
\end{figure} 

\begin{figure}[h]
   \setlength{\abovecaptionskip}{-5pt}
	\setlength{\belowcaptionskip}{5pt}
	\centering
	\includegraphics[width=3.6 in]{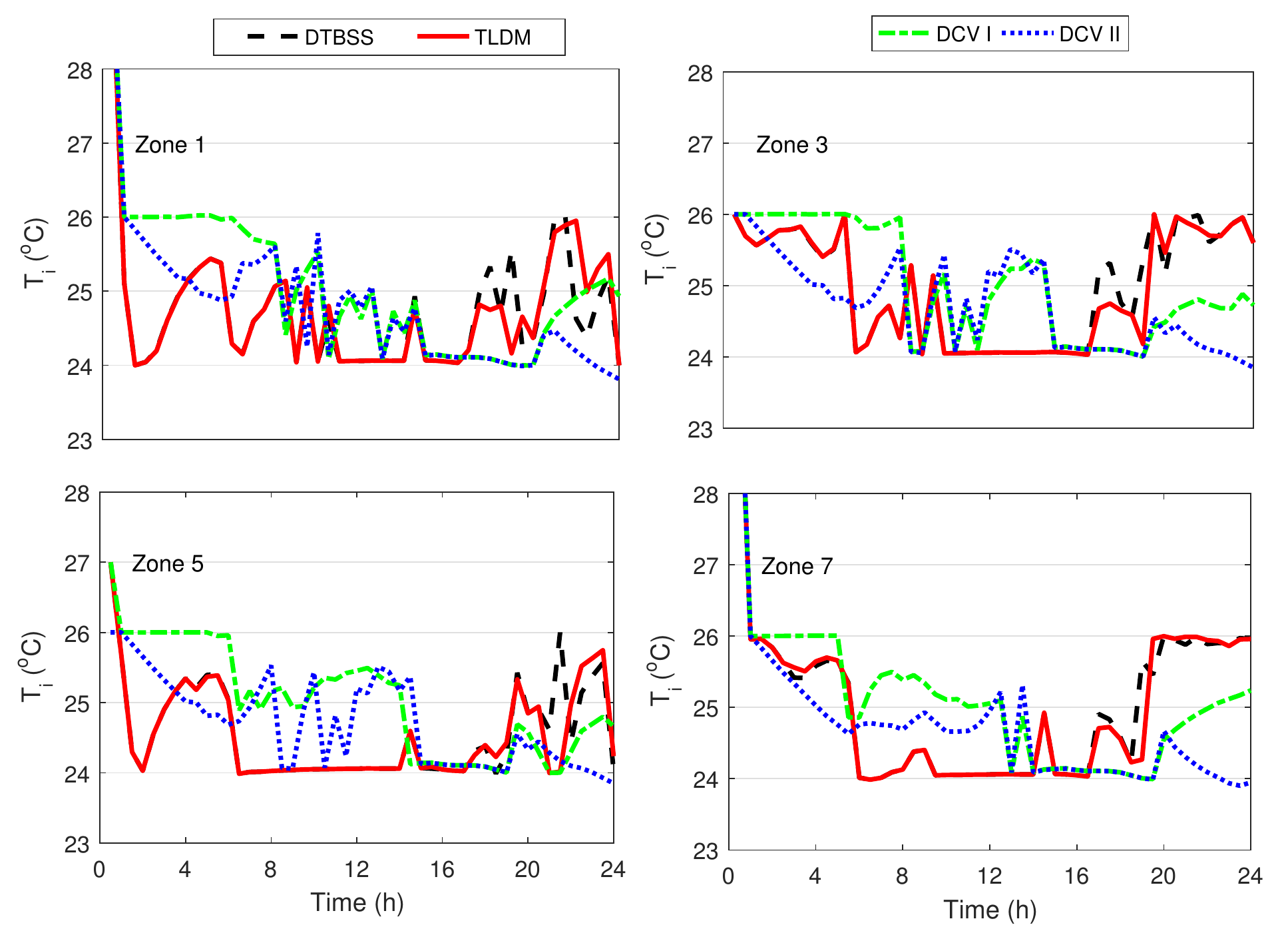}\\
	\caption{The temperature for the four zones among $I=50$ zones.} \label{zone temperature50}
\end{figure} 

\begin{figure}[h]
   \setlength{\abovecaptionskip}{-5pt}
	\setlength{\belowcaptionskip}{5pt}
	\centering
	\includegraphics[width=3.6 in]{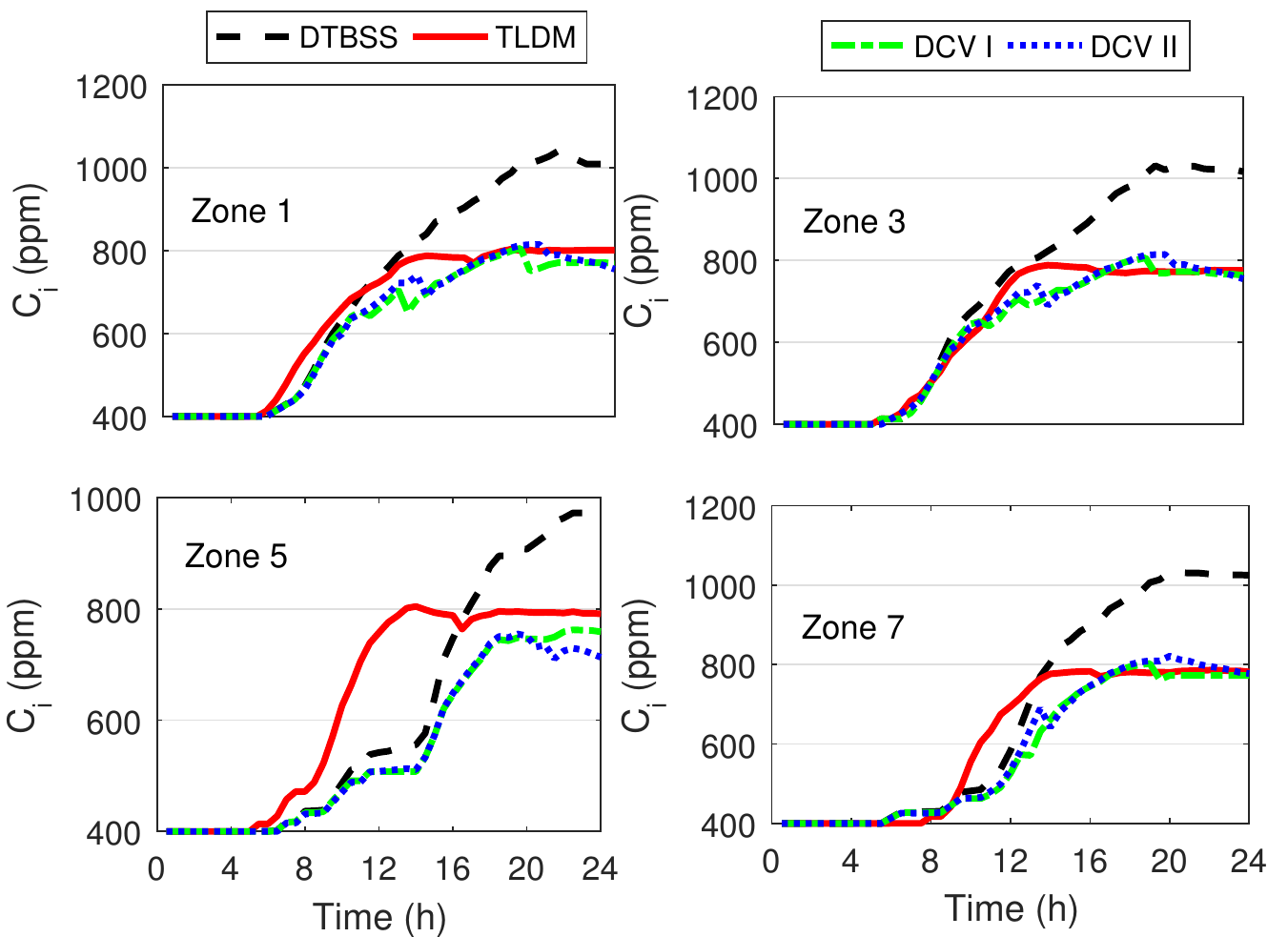}\\
	\caption{The CO$_2$ concentration  for the  four zones among $I=50$ zones} \label{zone CO250}
\end{figure} 

\begin{figure}[h]
    \setlength{\abovecaptionskip}{-5pt}
	\setlength{\belowcaptionskip}{5pt}
	\centering
	\includegraphics[width=3.2 in, height= 1.8 in]{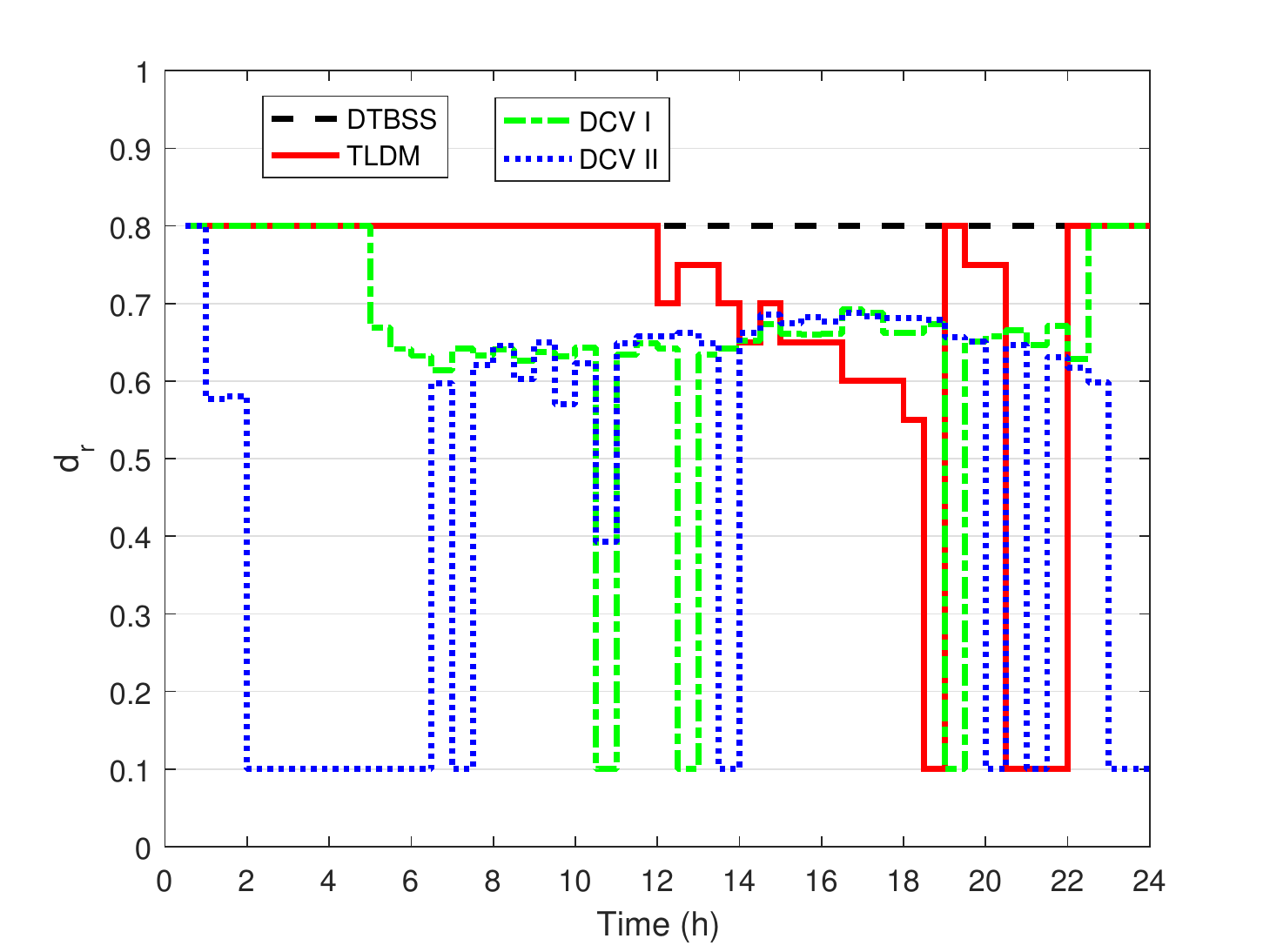}\\
	\caption{The ventilation rate  ($\bm{d}_r$) of the HVAC system} \label{dr50}
\end{figure}

As an instance, we further investigate the  50-zone case study.  
We display the zone  occupancy,  zone  mass flow rates,  zone temperature, and zone CO$_2$ for three randomly selected zones in  Fig. \ref{occupancy50}-\ref{zone CO250}.
From Fig. \ref{zone temperature50} -\ref{zone CO250}, we see  both the TLDM  and the DCVs maintain the comfortable zone temperature range ($[24, 26]^\circ$C) and zone CO$_2$ range ($[0, 800]$~ppm). Besides,  we observe close peak  (800~ppm) over the  zone CO$_2$ trajectories  under the TLDM and the DCVs, which implies the close IAQ maintained by these methods.
Besides, from Fig. \ref{zone temperature50},  we can have some insights in the  characteristics of the TLDM to achieve thermal comfort and IAQ. 
Specifically, we see the zone temperature approaches  the lower bound  during the working hours  with high  occupancy. 
This phenomenon is caused by the \emph{two-phase} structure  of  LLC in TLDM  where the  zone mass flow rates is first adjusted  to achieve IAQ and then  the ventilation rate is regulated  if necessary.

We also compare the ventilation rates ($\bm{d}_{r}$) of different methods in Fig. \ref{dr50}, which exhibit some interesting phenomenon to be interpreted.  First of all, we see the ventilation rate of TLDM corresponds well to the occupancy. 
Specifically,   we see relatively low ventilation rate  (larger $\bm{d}_{r}$)  during  the off-working  hours  but the opposite over the working hours (smaller $\bm{d}_{r}$). This is reasonable as the LLC is only  invoked to regulate the ventilation rate  if  the CO$_2$ is to be violated, which generally results from high occupancy. 
Surprisingly,  the results for  DCV  I and DCV II are almost opposite and quite different from the TLDM. 
Specifically,  we see  that the ventilation rate ($\bm{d}_{r}$) of DCV I corresponds well to the occupancy whereas the  DCV II presents the opposite. 
These phenomenon are  attributed to the rules that  used to determine zone fresh air flow rates  in the DCVs  as discussed before.  For DCV I, the zone fresh air totally depends on zone occupancy and thus we observe  a 
synchronous pace of ventilation rate ($\bm{d}_{r}$) with the  occupancy.  However, for DCV II, the zone mass flow rates are jointly determined by the occupancy and space. 
Therefore, during the off-working hours with low occupancy, we can observe higher ventilation rate  as the proportions of  zone fresh air flow rates  dominate in the total  zone mass flows rates (higher proportion) for maintaining  thermal comfort.

\section{Conclusion}
This paper studied scalable  control  of multi-zone  HVAC systems with the objective to reduce the energy cost for maintaining thermal comfort and IAQ simultaneously.  
This problem is  computationally challenging  due to the complex  system dynamics.
To cope with the  difficulties,  we proposed  a two-level distributed method (TLDM) which integrates  the \emph{upper}  and \emph{lower} level control by exploiting  the problem structures.  
Specifically, the \emph{upper} level  first optimizes  zone mass flow rates to satisfy  thermal comfort  with minimal energy cost and   the \emph{lower} level  regulates  zone mass flow rates and the ventilation rate   to achieve IAQ while preserving the \emph{near} energy saving performance of ULC.
As both the \emph{upper} and \emph{lower}  level control use  distributed computation, the proposed method is computationally efficient and scalable.
The  method's performance in energy saving and scalability  was demonstrated by  numeric studies. 
The sub-optimality of the proposed method  in energy cost is around  4\%  but with significant computational benefits over the centralized method.
Compared to the distributed Token-based scheduling strategy (DTBSS),   the proposed method induces a marginal increase of energy costs but provides IAQ.
In addition, the proposed method provided 8-10\% energy savings over the demand controlled ventilation strategies (DCVs).




%
\bibliographystyle{ieeetr}
\bibliography{reference}

\end{document}